\documentclass[twocolumn,showpacs,prb]{revtex4-2}
\usepackage{amsmath}
\usepackage{amssymb}
\usepackage{graphicx}
\usepackage{color}
\usepackage{multirow}
\usepackage{subfigure}

\begin{document}

\title{Nonlocal optical conductivity of Fermi surface nesting materials}

\author{Xiamin Huang$^{1}$, Xiao Jiang$^{2}$, Bing Huang$^{2}$ and Zhou Li$^{1,3}$}

\affiliation{$^{1}$ Guangdong Aerospace Information Research Institute, Guangzhou 510535, China}

\affiliation{$^{2}$ Beijing Computational Science Research Center, Beijing 100193, China}

\affiliation{$^{3}$ University of Chinese Academy of Sciences, Beijing 100039, China}

\providecommand{\tabularnewline}{\\}

\begin{abstract}
We investigate the nonlocal optical conductivity of Fermi surface nesting materials which support charge density waves or spin density waves. The nonlocal optical conductivity contains information of correlations in electron fluids which could not be accessed by standard optical probes. Half metal emerges from doping a charge density wave and similarly spin-valley half metal emerges from doping a spin density wave. Based on the parabolic band approximation, we find the Drude peak is shifted to higher frequency and splits into two peaks in the nonlocal optical conductivity. We attribute this to the two Fermi velocities in the half-metal or spin-valley half metal states.
\end{abstract}

\date{\today }

\maketitle

\section{Introduction}
The scattering-type scanning near-field optical microscope (s-SNOM) is frequently used to probe the response pattern of a specific material (e.g. graphene) \cite{Fei1,Fei2,Lundeberg} driven by electromagnetic waves oscillating in both time and space. The spatial dispersion in the response pattern is characterized by the \textit {nonlocal} optical conductivity in close connection to the complicated structure of Fermi surface and electron correlations \cite{book}. These short-range effects are washed out by the standard far-field optical probe, which \textit{only} measures the optical conductivity in the long wavelength limit. The tip radius $a$ is about 25 nm \cite{Fei1,Fei2,Lundeberg} for the atomic force microscope operating in the tapping mode of s-SNOM. Thus the wave vector $q$ is in the order of $1/a=0.04\mathrm{nm}^{-1}$.

The nonlocal optical conductivity is crucial for plasmons (collective oscillations of electrons). The imaginary part of the nonlocal optical conductivity determines the dispersion of plasmons, while the real part determines the damping rate of plasmons. Plasmonic antennas, lenses and resonators are successful \cite{Schuller} at concentrating electromagnetic energy into the subwavelength and deep-subwavelength volume. Density-independent plasmons were found in nodal-line semimetals \cite{WangJF}, with possible applications in terahertz-stable topological metamaterials.

While a weak electron-phonon interaction is one of the basic assumptions in the BCS theory \cite{BCS} for conventional superconductivity, strong electron-phonon interaction (EPI) leads to interesting phenomena such as polarons \cite{polaron0,polaron1} and charge density waves (CDW) \cite{Gruner, Monceau,Zhang,Lu}. The interplay of spin degree of freedom and strong EPI leads to spin polarons \cite{polaron2} and spin density waves (SDW) \cite{SDW1,SDW2}. One prevailing understanding of charge density waves is the Peierls instability \cite{Peierls} in one dimension, and the Fermi surface nesting in any dimension. In many real materials, it is realized that the Peierls instability fails and the CDW phase is determined from the momentum dependence of the electron-phonon coupling matrix element \cite{Plummer}.

Reference \cite{Rozhkov} pointed out that a spin-valley half-metal state emerges from doping a spin density wave state. Then reference \cite{Rakhmanov} clarified that weak repulsive electron-elsectron interaction contributes to the spin density wave ordering, following a BCS-like mean field approach. The Fermi surface in the half metal state is fully spin polarized. The nesting vector connecting particle and hole valleys gives inter-valley scattering. In principle, if no electron-phonon interaction is considered, the SDW order is always energy favored than the CDW. In the presence of electron-phonon interaction, one needs to consider the spin-flip process when the electrons are scattered by the lattice. The four-band Hamiltonian for the spin-valley half metal has also been applied in twisted bilayer graphene \cite{Tabert, Sboychakov}.

In two dimensional valleytronic materials such as $\rm{MoS_2}$ \cite{Xiao,Zhou} the two nonequivalent valleys are separated in the Brillouin zone by a large momentum so the inter-valley scattering is very small. Just like manipulating spin has lead to spintronics \cite{Wolf,Fabian}, manipulating valley index can produce new effects including using it to carry information. As an example, in the context of graphene, the reference \cite{Xiao1} showed that a contrasting intrinsic magnetic moment and Berry curvature are associated with the carrier valley index. The inter-band optical Hall conductivity \cite{Zhou} is then shown to be connected to the Berry curvature of the material, adding a dynamical factor to the static Berry curvature. 

The intra-band longitudinal optical conductivity has been less investigated. In the spin-valley half metal (doping form SDW) or half metal (doping form CDW), the spin and valley polarized Fermi surface and the Fermi velocities are however closely connected to the intra-band nonlocal optical conductivity. Recently, in transition metal dichalcogenides (such as $\rm{VSe_2}$, $\rm{MoTe_2}$, $\rm{WTe_2}$ and electron-doped $\rm{MoS_2}$ \cite{Feng,Chen,Calandra,Moha}) CDW phases were found. Half metals are predicted in two dimenional metal selenides such as V doped $\rm{SnSe_2}$ and $\rm{Co_2Se_3}$ \cite{CoSe}. In g-$\rm{C_3N_4}$ and g-$\rm{C_4N_3}$, the electrons of which are entirely from s and p orbitals, metal-free half-metals are predicted \cite{CN,CN1}. They may be used in bio-compatible applications, e.g. in synergistic photo-thermal treatment to kill cancer cells \cite{Bio}.

In this work, we present density-functional theory(DFT) calculations of Fermi surface nesting candidate materials, 1T-$\rm{VSe_2}$ and g-$\rm{C_4N_3}$, and provide an approximate tight binding Hamiltonian for 1T-$\rm{VSe_2}$. For the intra-band nonlocal optical conductivity, the anistropy of the band structure averages out, so we choose an isotropic parabolic band Hamiltonian to capture the important physics. We then use the Green's function to calculate the nonlocal optical conductivity based on a velocity-velocity correlation function approach. Both intra-band and inter-band optical conductivity are obtained, in one, two and three dimensions. The nonlocal ($q\ne0$) optical conductivity becomes anistropic in two and three dimensions, a feature not available in the ($q=0$) optical conductivity. The imaginary part of the nonlocal optical conductivity shows two peaks in the intra-band region for some parameters of the Fermi-surface nesting model, suggesting two branches of plasmons in the intra-band region.

\section{Fermi surface nesting candidate materials}

The numerical calculations on the mechanical and electronic properties of $\rm{VSe_2}$ were performed by using a first-principles method based on the DFT\cite{Kresse}, as implemented in the DS-PAW which is a program under the Device Studio platform. The DS-PAW is based on the plane wave basis and the projector augmented wave (PAW) representation\cite{Blochl}. The Perdew-Burke-Ernzerhof (PBE) exchange-correlation energy functional within the generalized gradient approximation (GGA) were employed\cite{Perdew}. In the calculations, atomic positions are fully relaxed until the force on each atom is less than $0.05 eV/\AA$. The electronic iteration convergence criterion is set to $10^{-4} eV$. The wave functions were expanded in plane waves up to a kinetic energy cutoff of $520 eV$. The Brillourin zone integration is obtained by using a k-point sampling mesh of $13\times13\times1$, generated according to the Gamma-centered method. In the standard DFT calculations which is used in this section, the staggered Hartree term representing the spin off-diagonal exchange energy is not included \cite{Gap}. In the future, we plan to develop atomistic simulation for SDW and CDW, similar to DFT based on Migdal-Eliashberg formalism \cite{Wudh}.

\begin{figure}
    \centering
    \subfigure [Crystal structure of 1T-$\rm{VSe_2}$.]{
    \includegraphics[width=3.2in]{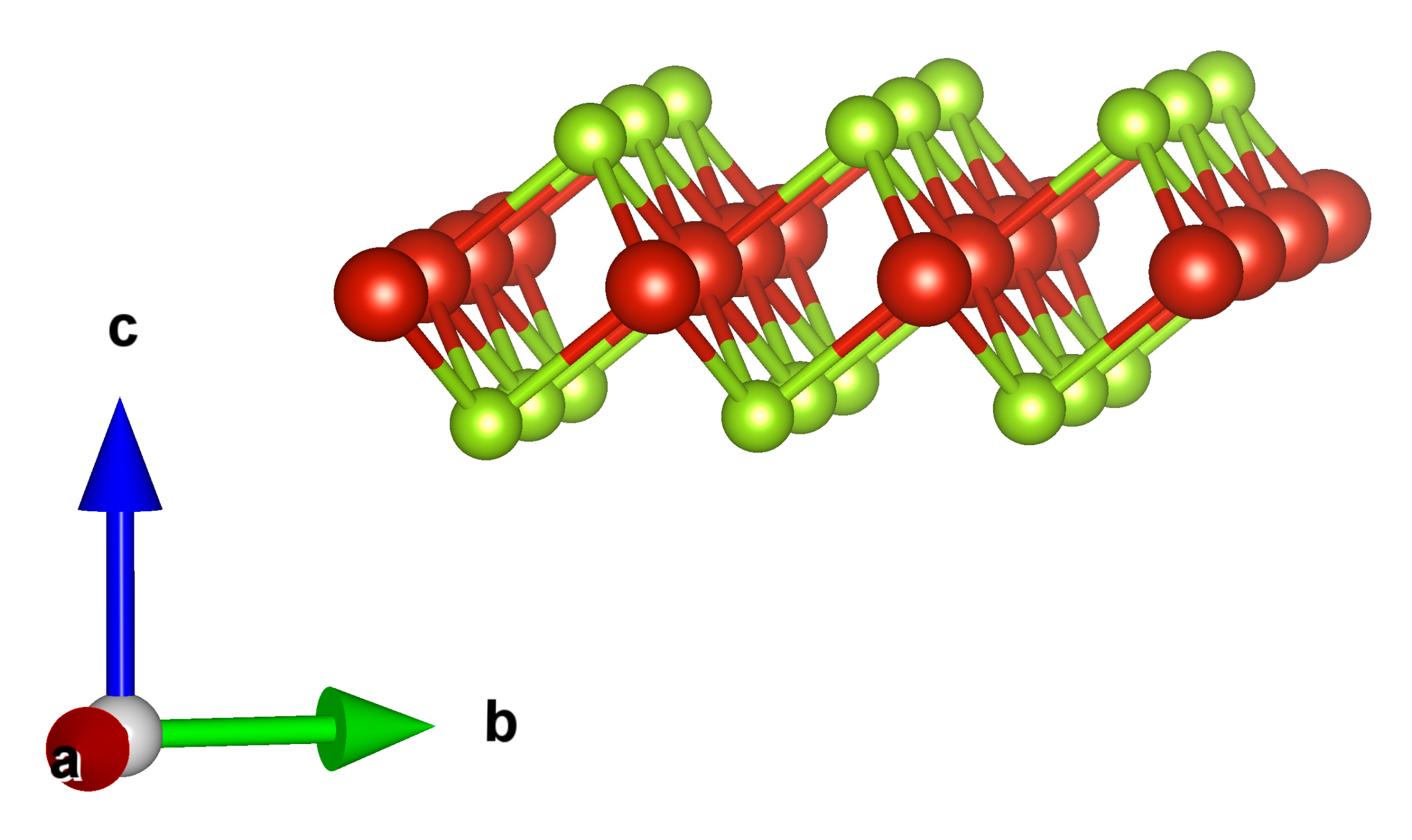}
    \label{fig:structure}}
    \quad
    \subfigure [The hopping between nearest and next-nearest atoms.]{
    \includegraphics[width=3.2in]{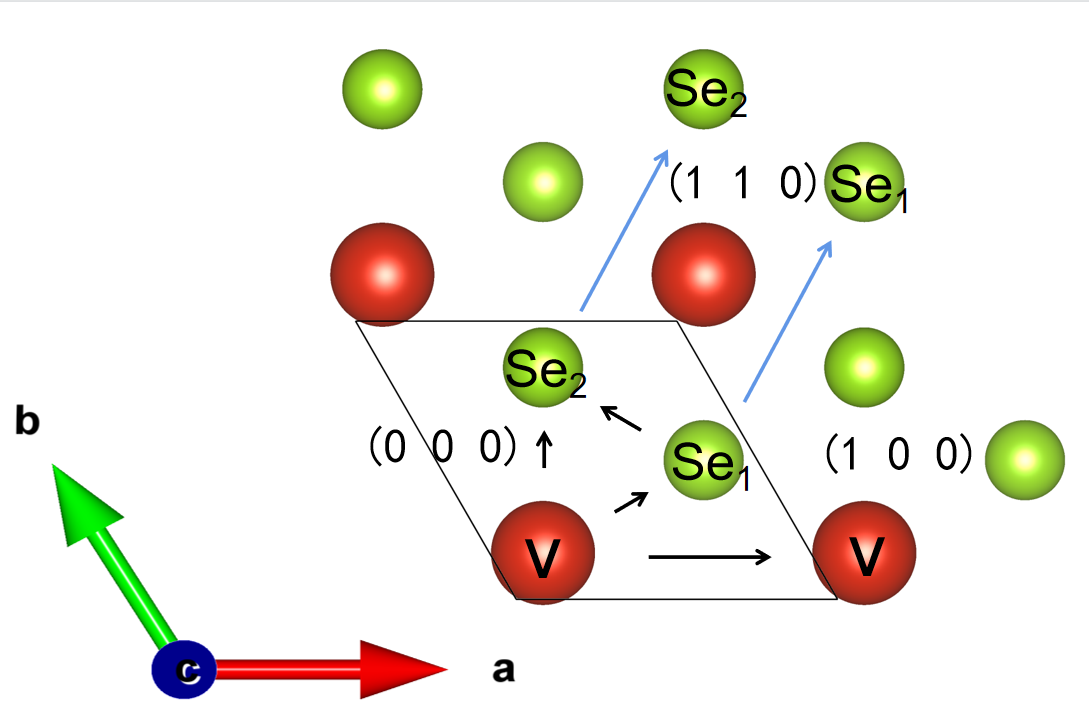}
    \label{fig:hopping}}
    \quad
    \caption{(Color online) Red balls represent V atoms; Green balls represent Se atoms. A unit cell with one V atom and two Se atom is marked in the bottom frame.}
    \label{fig:struct}
\end{figure}

\begin{figure}[tp]
    \begin{centering}
    \includegraphics[width=3.5in]{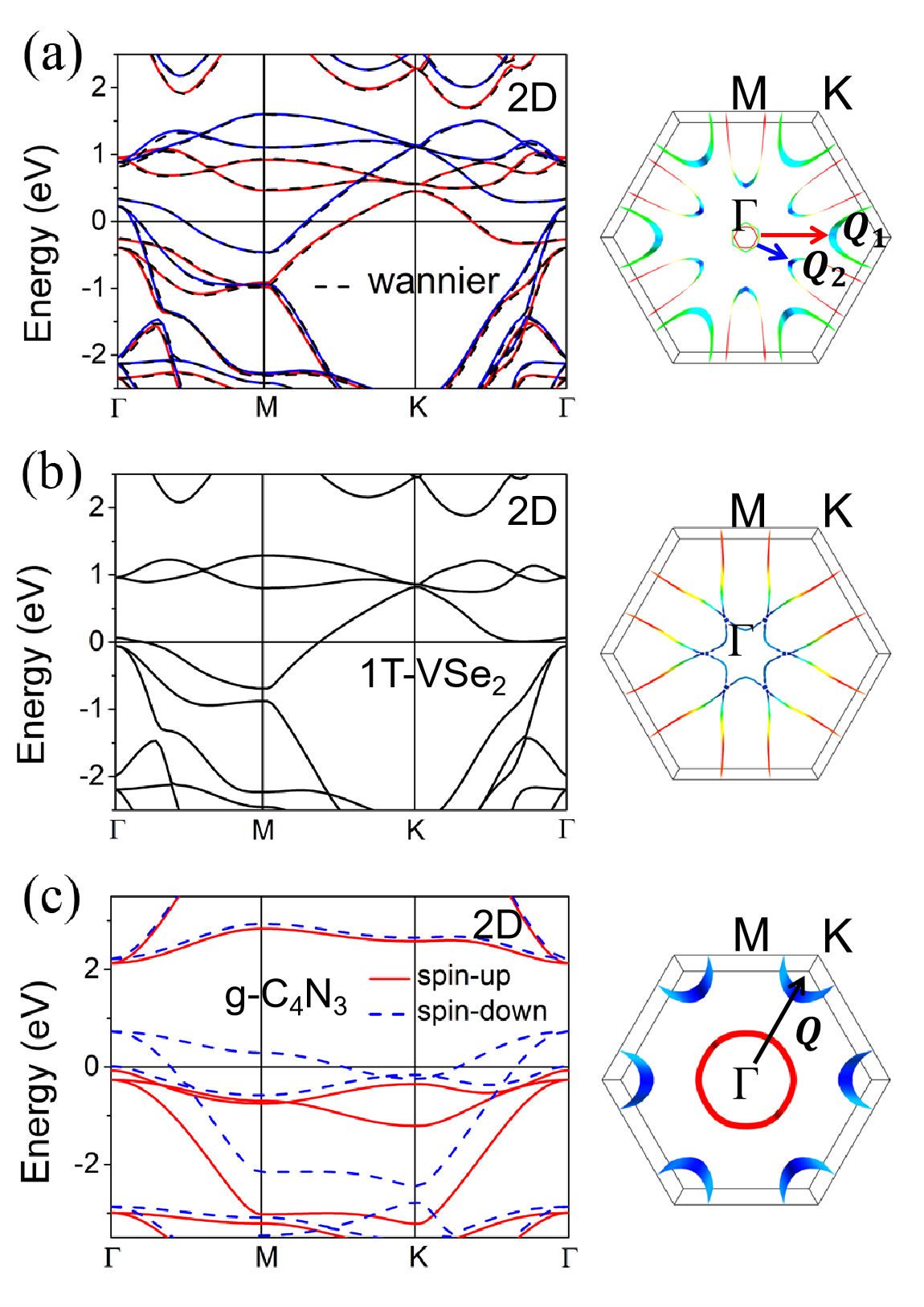}
    \par\end{centering}
    \caption{(Color online) The band-struture and Fermi surface of (a) spin-polarized DFT of monolayer 1T-$\rm{VSe_2}$, (b) non-spin-polarized DFT of monolayer 1T-$\rm{VSe_2}$ and (c) spin-polarized DFT of g-$\rm{C_4N_3}$. Red and blue lines in (a) and (c) represent the contributions from the spin-up and spin-down electrons, respectively, and the black dashed line in (a) represents the wannier90 fitting result. All materials are 2D. The Fermi energies of (a), (b) and (c) are -2.00 eV, -2.06 eV, -3.95 eV respectively. We then shift the whole band-structure to set the Fermi energies at 0 eV. In the right frames of (a), (b) and (c), we plot the Fermi surface (energy contours at 0 eV) in the 2D momentum Brillouin zone. We mark the nesting vector $\mathbf{Q}$ which connects different parts of the Fermi surface. }
    \label{fig:bandall}
\end{figure}

\begin{table}[]
\caption{The Se1-Se2 hopping amplitude.}
\begin{tabular}{|cc|ccc|}
\hline
\multicolumn{2}{|c|}{\multirow{2}{*}{}} & \multicolumn{3}{c|}{\textbf{Se2}} \\ \cline{3-5} 
\multicolumn{2}{|c|}{} & \multicolumn{1}{c|}{$p_y$} & \multicolumn{1}{c|}{$p_z$} & $p_x$ \\ \hline
\multicolumn{1}{|c|}{\multirow{3}{*}{\textbf{Se1}}} & $p_y$ & \multicolumn{1}{c|}{0.464} & \multicolumn{1}{c|}{-0.316} & 0.186 \\ \cline{2-5} 
\multicolumn{1}{|c|}{} & $p_z$ & \multicolumn{1}{c|}{-0.316} & \multicolumn{1}{c|}{0.126} & -0.089 \\ \cline{2-5} 
\multicolumn{1}{|c|}{} & $p_x$ & \multicolumn{1}{c|}{0.186} & \multicolumn{1}{c|}{-0.089} & 0.036 \\ \hline
\end{tabular}
\label{table:Se1-Se2}
\end{table}

\begin{table}[]
\caption{The V-Se1 and V-Se2 hopping amplitude.}
\begin{tabular}{|cc|ccc|ccc|}
\hline
\multicolumn{2}{|c|}{\multirow{2}{*}{}} & \multicolumn{3}{c|}{\textbf{Se1}} & \multicolumn{3}{c|}{\textbf{Se2}} \\ \cline{3-8} 
\multicolumn{2}{|c|}{} & \multicolumn{1}{c|}{$p_y$} & \multicolumn{1}{c|}{$p_z$} & $p_x$ & \multicolumn{1}{c|}{$p_y$} & \multicolumn{1}{c|}{$p_z$} & $p_x$ \\ \hline
\multicolumn{1}{|c|}{\multirow{5}{*}{\textbf{V}}} & $d_{xy}$ & \multicolumn{1}{c|}{0.515} & \multicolumn{1}{c|}{0.481} & 0.283 & \multicolumn{1}{c|}{-0.518} & \multicolumn{1}{c|}{0} & 0.568 \\ \cline{2-8} 
\multicolumn{1}{|c|}{} & $d_{yz}$ & \multicolumn{1}{c|}{0.487} & \multicolumn{1}{c|}{-0.750} & -0.729 & \multicolumn{1}{c|}{0} & \multicolumn{1}{c|}{-0.500} & 0 \\ \cline{2-8} 
\multicolumn{1}{|c|}{} & $d_{z^2}$ & \multicolumn{1}{c|}{0.278} & \multicolumn{1}{c|}{-0.738} & 0.095 & \multicolumn{1}{c|}{0.554} & \multicolumn{1}{c|}{0} & 1.160 \\ \cline{2-8} 
\multicolumn{1}{|c|}{} & $d_{xz}$ & \multicolumn{1}{c|}{-0.379} & \multicolumn{1}{c|}{-0.139} & 0.624 & \multicolumn{1}{c|}{-0.761} & \multicolumn{1}{c|}{0} & -0.389 \\ \cline{2-8} 
\multicolumn{1}{|c|}{} & $d_{x^2-y^2}$ & \multicolumn{1}{c|}{0.655} & \multicolumn{1}{c|}{0.464} & -0.148 & \multicolumn{1}{c|}{0} & \multicolumn{1}{c|}{-0.696} & 0 \\ \hline
\end{tabular}
\label{table:V-Se1Se2}
\end{table}
\begin{table}[]
\caption{The Se-Se next nearest hopping amplitude.}
\begin{tabular}{|l|l|l|l|l|l|}
\hline
 & $p_y$,$p_y$ & $p_z$,$p_z$ & $p_z$,$p_x$ & $p_x$,$p_z$ & $p_x$,$p_x$ \\ \hline
\textbf{Se1(000)\textgreater{}Se1(110)} & -0.169 & 0.193 & 0.344 & 0.553 & 0.654 \\ \hline
\textbf{Se2(000)\textgreater{}Se2(110)} & -0.169 & 0.193 & 0.553 & 0.344 & 0.654 \\ \hline
\end{tabular}
\label{table:next_nearest}
\end{table}

In Fig.~\ref{fig:struct}(a), we give a schematic of the structure of 1T-$\rm{VSe_2}$ in the trigonal phase corresponding to the space group $P-3m1$. For a single-layer 1T-$\rm{VSe_2}$ the primitive cell contains three atoms, and the optimized lattice constant is $a = b = 3.33 \AA$, and the angle between them is $\gamma=120^\circ$.
In Fig.~\ref{fig:struct}(b), we show a super-cell of $\rm{VSe_2}$, and mark the three unit cells (000), (100), and (110). The black arrows represent the hopping between the nearest neighbor atoms, including the Se1-Se2, V-Se1, V-Se2 hoppings in the unit cell (000), and the V-V hopping between the unit cell (000) and unit cell (100). The blue arrows represent the second-nearest neighbor hopping, including Se1-Se1 and Se2-Se2 hoppings between the unit cell (000) and unit cell (110).

In table I, II and III, we list the Wannier fitting results of the Se-Se and V-Se hopping parameters. The V-V hopping parameters is small, however the next nearest Se-Se hopping parameter is large. The fitted tight binding model is a very large Hamiltonian matrix, to be manipulated numerically. From table I, II and III, we provide an approximate tight binding model, as given in the supplementary material. The anistropy of the tight binding model averages out in the intra-band conductivity, so we use the isotropic parabolic band approximation to capture important physics of the many-body calculations in the section III.

In Fig.~\ref{fig:bandall}, we show the band-structure and Fermi surface of 1T-$\rm{VSe_2}$ and g-$\rm{C_4N_3}$. In Fig.~\ref{fig:bandall}(a), the spin-polarized calculations predicts monolayer 1T-$\rm{VSe_2}$ is a magnetic metal, of which the magnitude of the magnetic moment is 0.6$\mathrm{\mu B}$/cell (DFT result). The Wannier fitting of DFT band structure is implemented within the WANNIER90 code\cite{Marzari}. We construct the Hamiltonian in the Wannier basis using five d orbitals of V atom and six p orbitals of two Se atoms to generate the localized Wannier functions. Overall, the Wannier fitted band structure agrees well with the DFT one. Some experiment data from angle-resolved photo-emission spectroscopy (ARPES) of $\rm{VSe_2}$ agrees well with the non-spin-polarized DFT \cite{VSe2}, while other experiment data of the Fermi surface agrees well with the spin-polarized DFT \cite{VSe2F}, so we present both of them in Fig.~\ref{fig:bandall} (a) and (b).  For the bulk $\rm{VSe_2}$, the band-structure is similar to 2D $\rm{VSe_2}$ near the $\Gamma$, $M$, $K$ points, as given in the supplementary material. There are more high symmetry momentum points $H$, $A$, $L$ for the bulk $\rm{VSe_2}$, and the band structure around these points are different as expected. In Fig.~\ref{fig:bandall}(c), the spin-polarized calculations predicts that g-$\rm{C_4N_3}$ is a half-metal. The direct band gap for the spin-up electrons is 2.20 $eV$ and the magnetic moment of 1 $\mathrm{\mu B}$/cell is found for g-$\rm{C_4N_3}$. Note that in the Fermi surface of g-$\rm{C_4N_3}$, a nesting vector $\mathbf{Q}$ clearly connects the circle around the $\Gamma$ point and the K point. In all the calculations, the staggered Hartree term is not included. In the following sections, we use a parabolic band approximation to discuss the impact of the staggered Hartree term.

\section{Hamiltonian and nonlocal optical conductivity}
In below, we use the parabolic band approximation and discuss the many body calculation for the non-local optical conductivity in the case of Fermi surface nesting, in one, two and three dimensions. With a nesting vector $\mathbf{Q}$ to connect (part of) the Fermi surface of a conduction band to that of a valence band, the Hamiltonian for a charge density wave state is given by 
\begin{eqnarray}
\hat{H} & = & \left[\begin{array}{cccc}
\frac{k^{2}}{2m}-\varepsilon_{0} & 0 & 0 & \Delta_{\downarrow}\\
0 & \frac{k^{2}}{2m}-\varepsilon_{0} & \Delta_{\uparrow} & 0 \\
0 & \Delta_{\uparrow} & -\frac{k^{2}}{2m}+\varepsilon_{0} & 0\\
\Delta_{\downarrow} & 0 & 0 & -\frac{k^{2}}{2m}+\varepsilon_{0}
\end{array}\right]\label{H0}
\end{eqnarray}
where $\Delta_{\uparrow}$ and $\Delta_{\downarrow}$ are the gap
parameters for a charge density wave insulator. In this paper we assume the gap parameters are real numbers so the Hamiltonian [Eq.~(\ref{H0})] is a hermitian matrix. In the case of complex gap parameters, two of the $\Delta_{\uparrow}$ and $\Delta_{\downarrow}$ in Eq.~(\ref{H0}) should be replaced by  $\Delta_{\uparrow}^*$ and $\Delta_{\downarrow}^*$ to keep it a hermitian matrix. The basis states for the Hamiltonian are $[c^{\dag}_{a,\mathbf{k},\downarrow},c^{\dag}_{a,\mathbf{k},\uparrow},c^{\dag}_{b,\mathbf{k}+\mathbf{Q},\uparrow},c^{\dag}_{b,\mathbf{k}+\mathbf{Q},\downarrow}]$; here $c^{\dag}_{a,\mathbf{k},\sigma}$ is the creation operator for a fermionic field with wave vector $\mathbf{k}$, spin $\sigma$ and valley $a$. $\sigma=\pm1$ is for the up and down direction of the spin. The spin density wave phase is connected to the charge density wave phase by a transformation $c^{\dag}_{b\uparrow}\rightarrow c^{\dag}_{b\downarrow}$  and $c^{\dag}_{b\downarrow}\rightarrow c^{\dag}_{b\uparrow}$ with the corresponding spin density wave order parameter defined as $\Delta_{\sigma}=<c^{\dag}_{a\sigma}c_{b\bar\sigma}>$ (here $\bar\sigma=-\sigma$). Although SDW and CDW are connected mathematically, in the experiment the detection is very different. X-Ray diffraction (XRD) and neutron-scattering is used to detect CDW, while only neutron-scattering is used to detect SDW. The gap parameters are generally determined by a variational principle based on the free energy of the system. For example, in \cite{Rozhkov} it was found $\Delta_{\uparrow}(x)=\Delta_{0}\sqrt{1-x/(\Delta_{0}N_{F})}$, $\Delta_{\downarrow}(x)=\Delta_{0}$ and $\mu=\Delta_{0}-x/2N_{F}$, where $x$ is the partial doping $x=-\partial\Omega/\partial\mu$, $\Omega$ is the grand potential and $\mu$ the chemical potential. The Green's function is defined as the inverse of $z-\hat{H}$, where $z=i\omega_n$ for imaginary frequency or $z=\omega+i\delta$ for real frequency,
\begin{eqnarray}
\hat{G}(\mathbf{k},z) & = & \frac{1}{z-\hat{H}}=\left[\begin{array}{cc}
G_{ee} & G_{eh}\\
G_{he} & G_{hh}
\end{array}\right]
\end{eqnarray}
We first define a velocity-velocity correlation function $\Pi_{xx}$ which depends on both the wave vector and the imaginary frequency, 
\begin{eqnarray}
&&\Pi_{xx}(\mathbf{q},i\omega_{n})= \notag\\ 
&&T\sum_{\mathbf{k},l}\mathrm{Tr}\langle\hat{v}_{x}\hat{G}(\mathbf{k}-\mathbf{q}/2,i\omega_{l})\hat{v}_{x}\hat{G}(\mathbf{k}+\mathbf{q}/2,i\omega_{l}+i\omega_{n})\rangle
\end{eqnarray}
here $\omega_{n}=2n\pi T$, $\omega_{l}=(2l+1)\pi T$ are the bosonic and fermionic Matsubara frequencies, $n$ and $l$ are integers. The nonlocal optical conductivity in real frequency is obtained from the Kubo formula by performing the analytical continuation $i\omega_{n}\rightarrow\omega+i\delta$,
\begin{equation}
\sigma_{xx}(\mathbf{q},\omega)=-\frac{e^{2}}{i\omega} \Pi_{xx}(\mathbf{q}, i\omega_{n}\rightarrow\omega+i\delta)
\end{equation}
The matrix Green's function $\hat{G}(\mathbf{k},z)$ can be expanded in terms of a matrix spectral density $\hat{A}(\mathbf{k},\omega)$ as 
\begin{equation}
\hat{G}(\mathbf{k},z)=\int_{-\infty}^{\infty}\frac{d\omega}{2\pi}\frac{\hat{A}(\mathbf{k},\omega)}{z-\omega}
\end{equation}
The matrix spectral density function is proportional to the imaginary part of the matrix Green's function, 
\begin{equation}
\hat{A}(\mathbf{k},\omega)=2\mathrm{Im}[\hat{G}(\mathbf{k},\omega)]\label{Green-1}
\end{equation}
After taking the sum over the imaginary frequencies $i\omega_{l}$, the longitudinal conductivity $\sigma_{xx}(\omega)$ is given by 
\begin{eqnarray}
\sigma_{xx}(\mathbf{q},\omega)=-\frac{e^{2}}{4\pi^{2}i\omega}\int_{-\infty}^{\infty}d\omega_{1}d\omega_{2}F(\omega)  \notag \\
 \sum_{\mathbf{k}}\mathrm{Tr}\langle\hat{v}_{x}\hat{A}(\mathbf{k}-\mathbf{q}/2,\omega_{1})\hat{v}_{x}\hat{A}(\mathbf{k}+\mathbf{q}/2,\omega_{2})\rangle \label{Sigma}
\end{eqnarray}
where $F(\omega)=\frac{[f(\omega_{1})-f(\omega_{2})]}{\omega-\omega_{2}+\omega_{1}+i\delta}$ and $f(x)$ is the Fermi-Dirac
distribution function defined as $f(x)=1/[\exp(x/T-\mu/T)+1]$. The real part of the optical conductivity can be obtained as 
\begin{eqnarray}
 \mathrm{Re}\sigma_{xx}(\mathbf{q},\omega)=\frac{e^{2}}{4\pi\omega}\int_{-\infty}^{\infty}d\omega_{1}[f(\omega_{1})-f(\omega_{1}+\omega)]\times \notag \\
\sum_{\mathbf{k}}\mathrm{Tr}\langle\hat{v}_{x}\hat{A}(\mathbf{k}-\mathbf{q}/2,\omega_{1})\hat{v}_{x}\hat{A}(\mathbf{k}+\mathbf{q}/2,\omega_{1}+\omega)\rangle \label{ReSi}
\end{eqnarray}
The spectral function Eq.~(\ref{Green-1}) is a delta-function if no impurity or other similar scattering are considered. Here we consider the self-energy correction from the impurity scattering to be a pure imaginary number, $i/\tau$. The spectral function Eq.~(\ref{Green-1}) is then a broadened delta-function and the integral over $\omega_1$ in  Eq.~(\ref{ReSi}) could not be removed. From Eq.~(\ref{Sigma}) and Eq.~(\ref{ReSi}) we write the imaginary part of the optical conductivity as
\begin{equation}
 \mathrm{Im}\sigma_{xx}(\mathbf{q},\omega)=\frac{1}{\pi}\int_{-\infty}^{\infty}d\omega'\frac{\mathrm{Re}\sigma_{xx}(\mathbf{q},\omega')}{\omega-\omega'}
\end{equation}
which recovers the Kramers-Kronig relation.

In one dimension the corresponding velocity operator can be obtained
as (for simplicity we set $\hbar=1$) 
\begin{eqnarray}
\hat{v}_{x}(\mathbf{k}) & =\frac{\partial\hat{H}}{\partial \mathbf{k}}= & \left[\begin{array}{cc}
(\mathbf{k}/m)\hat{I_{2}} & 0\\
0 & -(\mathbf{k}/m)\hat{I_{2}}
\end{array}\right]
\end{eqnarray}
where $I_{2}$ is the 2 by 2 unit matrix. For the CDW state, the spin operator is $S_z=\mathrm{diag}(-1,1,1,-1)$. For the SDW state, the spin operator is changed to $S_z=\mathrm{diag}(-1,1,-1,1)$, while the spin-valley operator is $S_z\tau_z=\mathrm{diag}(-1,1,1,-1)$, so the spin operator for CDW is the same as the spin-valley operator for SDW. For both cases the valley operator is $\tau_z=\mathrm{diag}(1,1,-1,-1)$. With these operators we define the spin conductivity (for a CDW), spin-valley conductivity (for a SDW) and valley conductivity (for both). Details are given in the supplementary material.

\begin{figure}[tp]
\begin{centering}
\includegraphics[width=3.2in,height=3in]{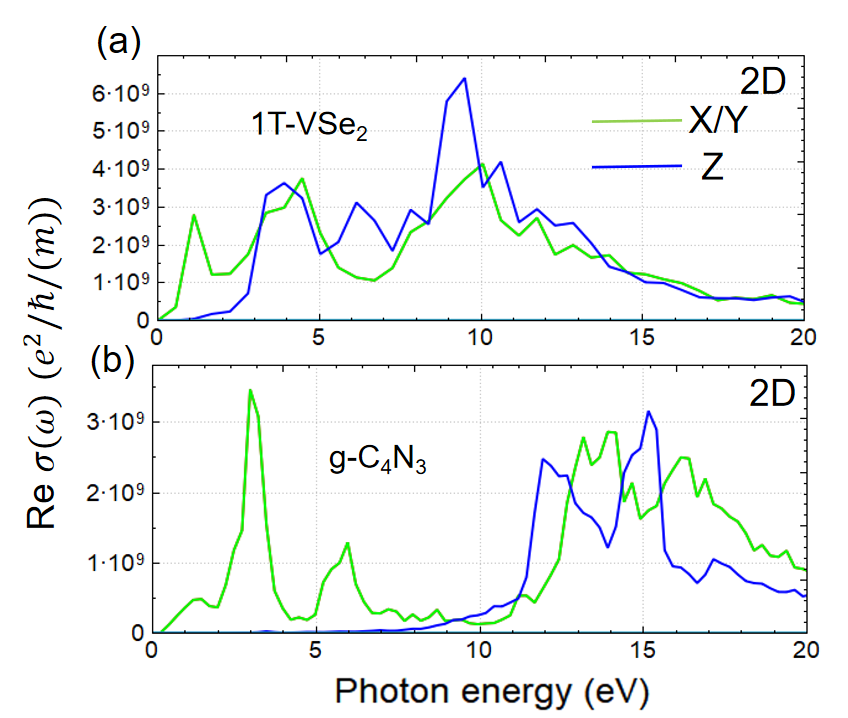} 
\end{centering}
\caption{(Color online) Real part of the optical conductivity at $\mathbf{q}=0$ for 2D materials 1T-$\rm{VSe_2}$ and g-$\rm{C_4N_3}$. In DFT calculations, the 2D material is treated as a 3D material with very large separation between 2D layers, so the unit of the optical conductivity is 3D. And we see an isotropic optical conductivity along x or y directions inside the 2D plane. The optical conductivity along the z direction outside the 2D plane is different. In the DFT software (DS-PAW) we use, the optical conductivity is calculated from the imaginary part of the dielectric function $\epsilon(\omega)$, given by Re$\sigma(\omega)=\omega\epsilon_0$Im$\epsilon(\omega)$, $\epsilon_0$ is the vacuum dielectric constant.}
\label{figdft} 
\end{figure}

\begin{figure}
\begin{centering}
 \subfigure{
\includegraphics[width=3.2in,height=3in]{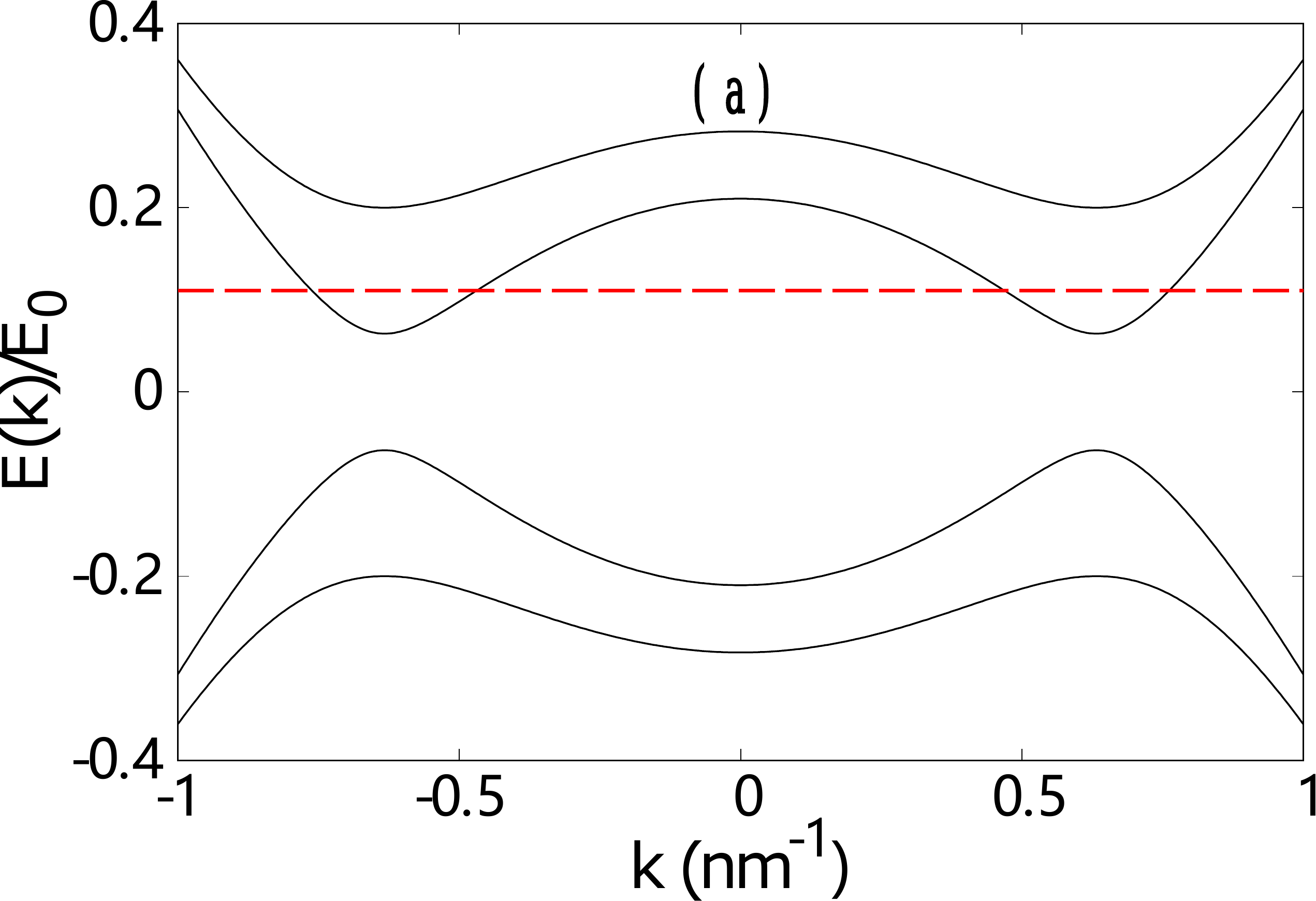} }
\subfigure{
\includegraphics[width=3.2in,height=3in]{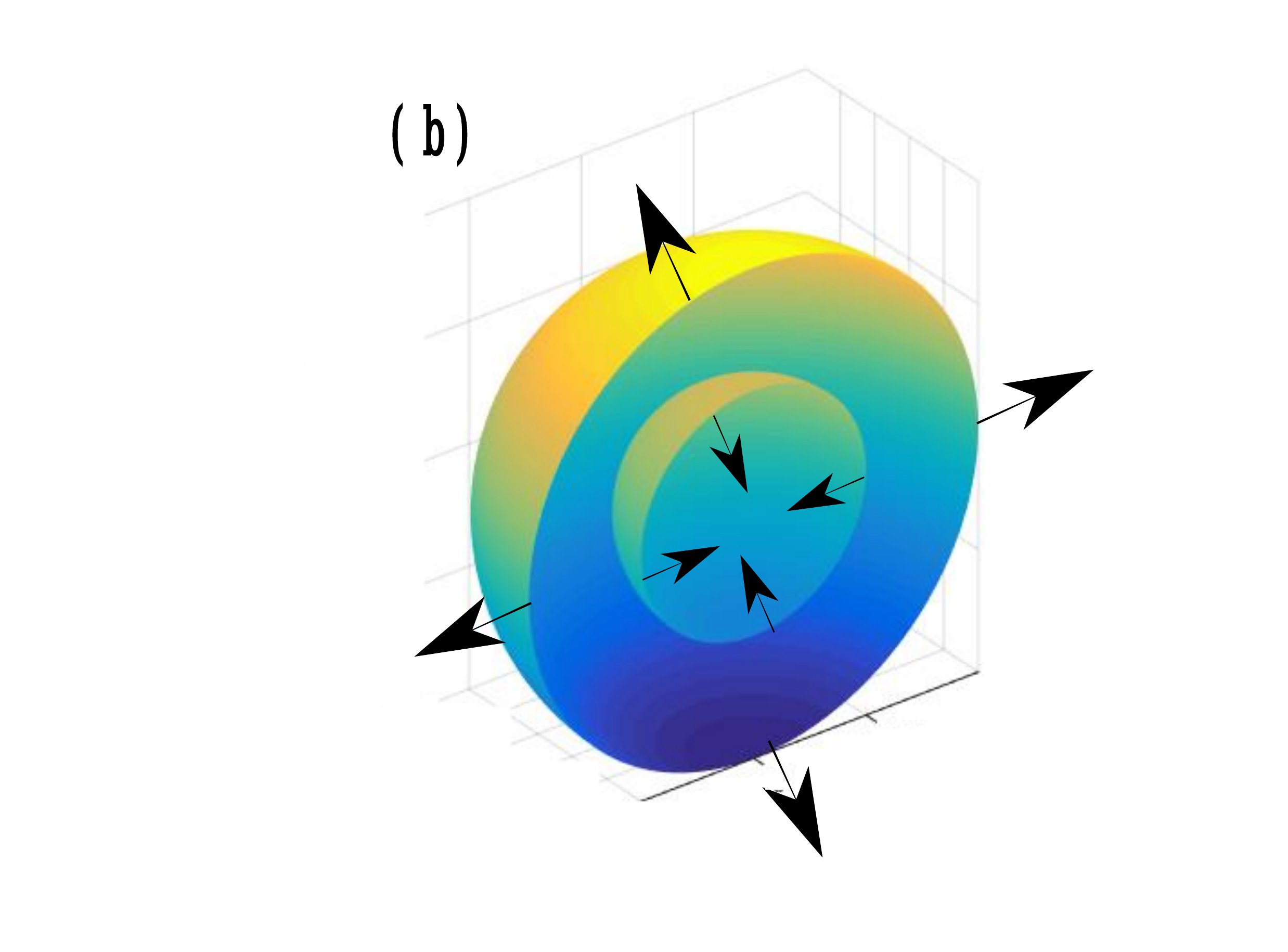} }
\end{centering}
\caption{(Color online) Top panel (a): Band structure for a Fermi surface nesting model in one dimension (1D) with the spin density wave order parameters $\Delta_{\uparrow}=0.063E_0$ and $\Delta_{\downarrow}=0.2E_0$. The chemical potential $\mu=0.11E_0$ is denoted with the red dashed line. Bottom panel (b): schematics of the iso-energy surface of a Fermi surface nesting model in three dimension (3D), we clearly see two Fermi-spheres assigned with two Fermi-velocities. When the chemical potential is tuned, the two Fermi-spheres moves close to or far away from each other, until one of them disappears. The Fermi-velocities point outward with the same amplitude on one Fermi-sphere, and point inward with smaller amplitude on the other, denoted by the black arrows. }
\label{Fig4} 
\end{figure}

\section{Results and discussions}

In Fig.~\ref{figdft}, we present the numerical results of the optical conductivity based on the DFT calculations in section II. In the DFT software (DS-PAW) we use, the real part of the optical conductivity is calculated from the imaginary part of the dielectric function. The DFT calculation is not able to catch the intra-band part of the nonlocal optical conductivity, usually in the frequency region from 0 to 50 meV. Also, in the standard DFT calculation, the many body effect from Fermi surface nesting is missing, because the staggered Hartree term is not included. In below, we present numerical results of the nonlocal optical conductivity based on the Hamiltonian [Eq.~(\ref{H0})] in section III.

The eigenvalues of the Hamiltonian [Eq.~(\ref{H0})] are obtained as $E=\pm\sqrt{\Big(\frac{k^{2}}{2m}-\varepsilon_{0}\Big)^2+\Delta^{2}}$, where $\Delta=\Delta_{\uparrow}$ or $\Delta_{\downarrow}$. We choose a typical wave-vector $k_0=1.0$ $\mathrm{nm}^{-1}$, the corresponding energy is $E_0=\hbar^{2}k_0^{2}/m_e=0.076$ $\mathrm{eV}$, where $m_e=9.1\times10^{-31}$ kg is the mass of an electron. For simplicity we set the effective mass $m=1$, the energy shift $\varepsilon_{0}=0.2 E_0$ and define $x'=x/(N_F\Delta_0)$ in all the numerical simulations. The impurity-scattering self-energy correction is $1/\tau=0.005E_0$. Note that the parameters of the energy shift $\varepsilon_{0}$, the gap ($\Delta_{\uparrow}$, $\Delta_{\downarrow}$) and the chemical potential $\mu$ are determined from a variational principle calculation \cite{Rozhkov}, it is suggested that the gap $\Delta_0$ is much smaller than $\varepsilon_{0}$, (e.g. $\Delta_0$=$\varepsilon_{0}$/4). Here we choose $\Delta_0=\varepsilon_{0}$. For the intra-band optical conductivity, tuning the band gap to be much smaller will not change the key feature, in the sense that the chemical potential still intersects with the lower conduction band and two Fermi velocities was found.

In Fig.~\ref{Fig4}, we show a typical 1D band structure (left) or a 3D iso-energy surface (right) for CDW/SDW phase. The doping parameter $x'=0.9$, so the gap parameters are $\Delta_{\uparrow}=0.063 E_0$ and $\Delta_{\downarrow}=0.2 E_0$. The doping $x$ gives an asymmetry between the gap order parameter for spin up and spin down, thus the Fermi surface is now spin polarized typical for a half-metal. The Fermi velocities are depicted by arrows in the right panel.  

In 1D, $q$ is a number, we only have the longitudinal conductivity, in 2D and 3D, $\mathbf{q}$ is a vector, we need to consider the longitudinal one ($\mathbf{q}$ along $\hat{e}_{x}$) and the transverse one \cite{Woods} ($\mathbf{q}$ perpendicular to $\hat{e}_{x}$), where $\hat{e}_{x}=(1,0)$ for 2D and $\hat{e}_{x}=(1,0,0)$ for 3D. For the longitudinal conductivity $\sigma^L_{xx}(q,\omega)$ we have the absolute value of
$\mathbf{k}+\mathbf{q}=\sqrt{(k_x+q)^2+k_y^2}=\sqrt{k^2+2kq\rm{cos}(\theta)+q^2}$.
For the transverse one $\sigma^T_{xx}(q,\omega)$ we have the absolute value of $\mathbf{k}+\mathbf{q}=\sqrt{(k_y+q)^2+k_x^2}=\sqrt{k^2+2kq\rm{sin}(\theta)+q^2}$. Because of this difference, we see a large anistropy in the nonlocal optical conductivity in two and three dimensions. 

\begin{figure}[tp]
\begin{centering}
\includegraphics[width=3.2in,height=3in]{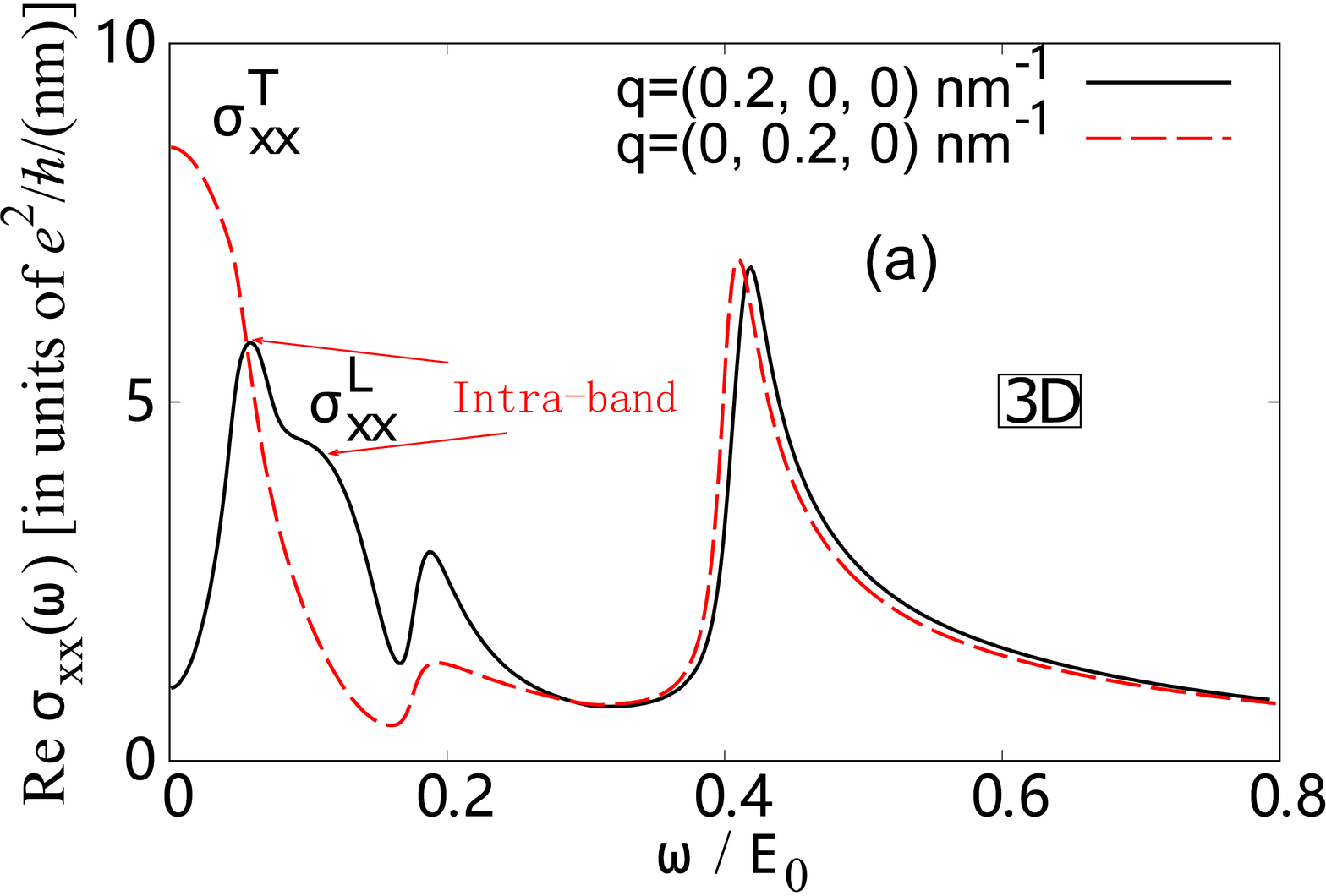} 
\includegraphics[width=3.2in,height=3in]{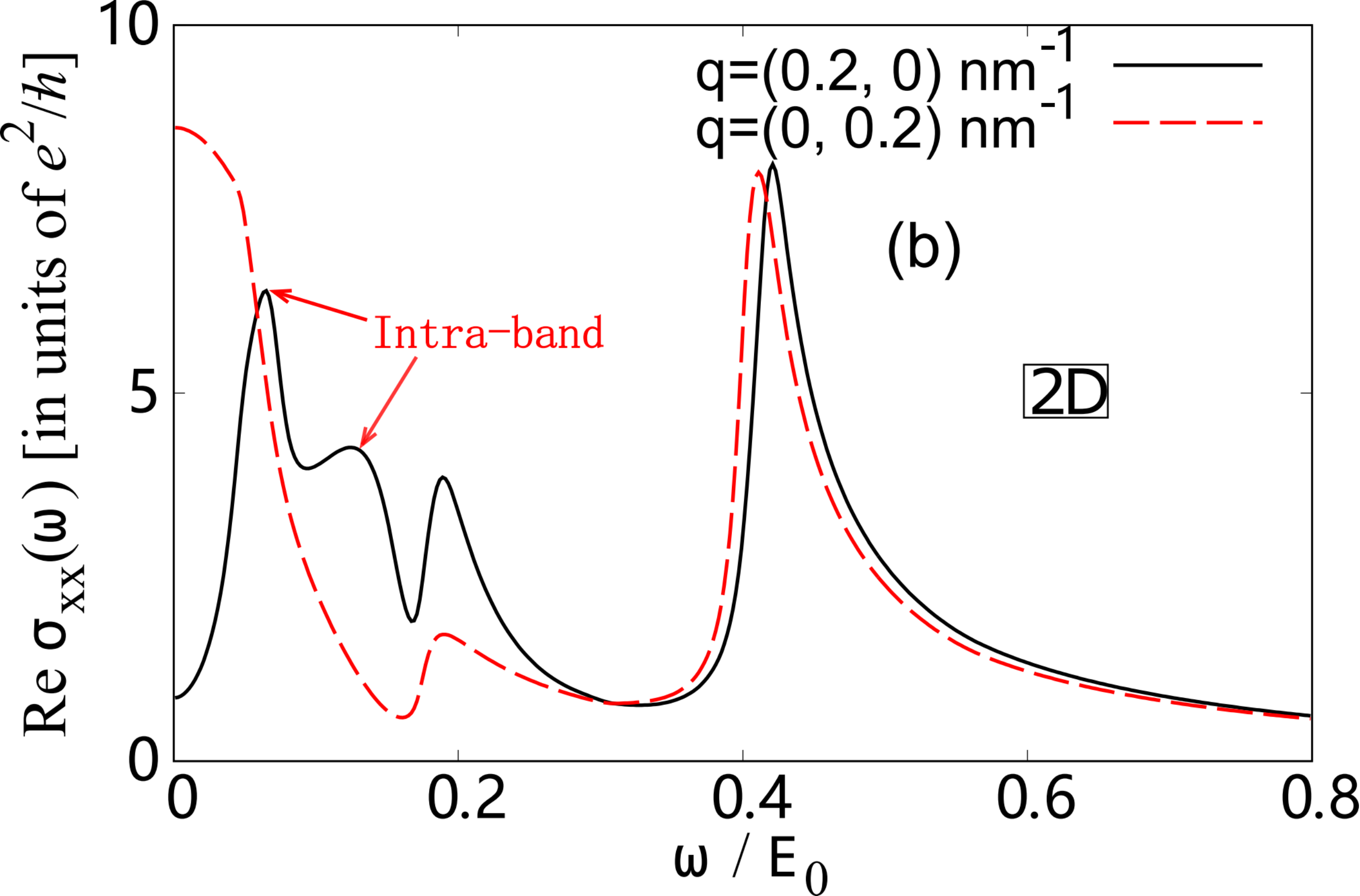} 
\includegraphics[width=3.2in,height=2.8in]{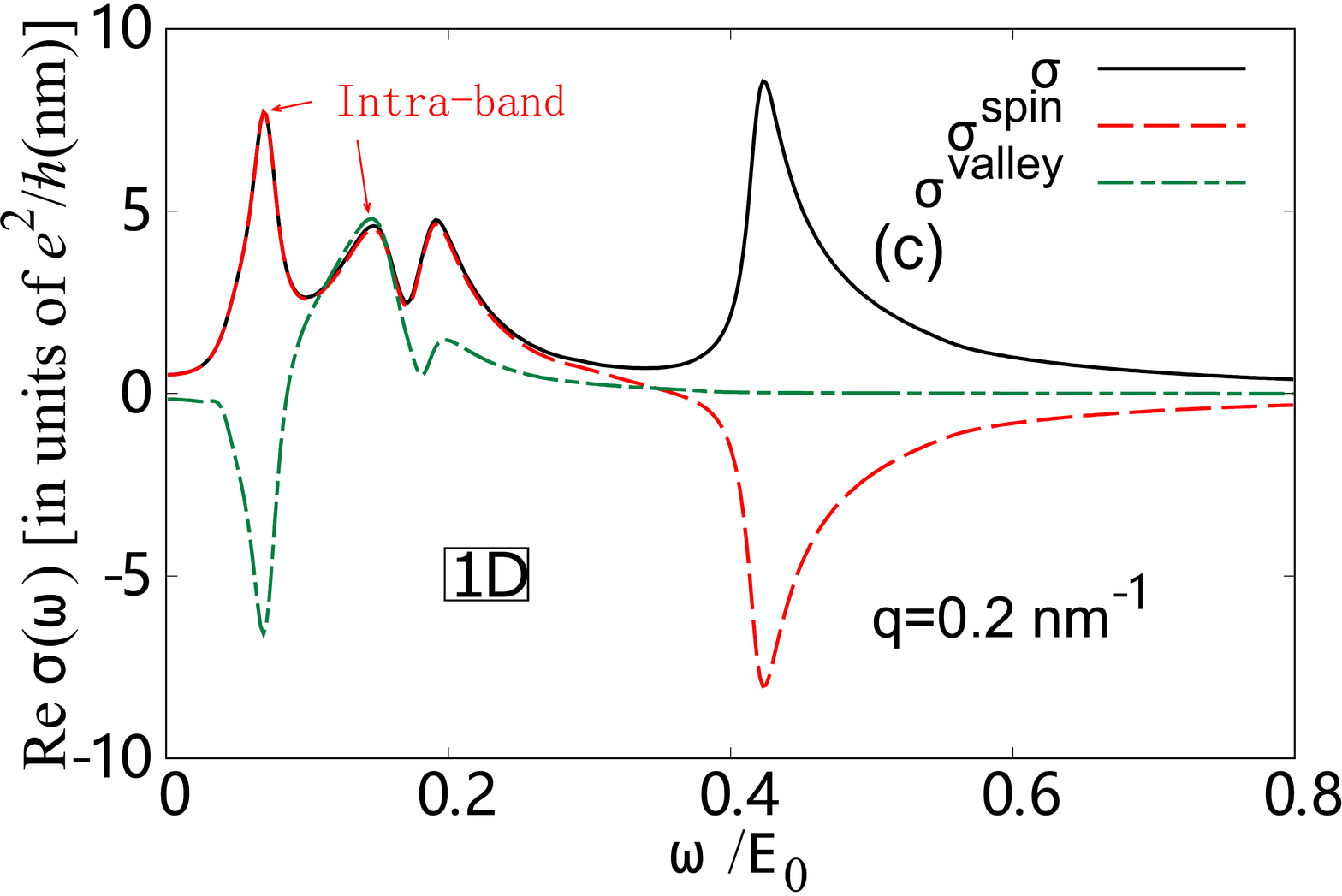} 
\includegraphics[width=3.2in,height=2.8in]{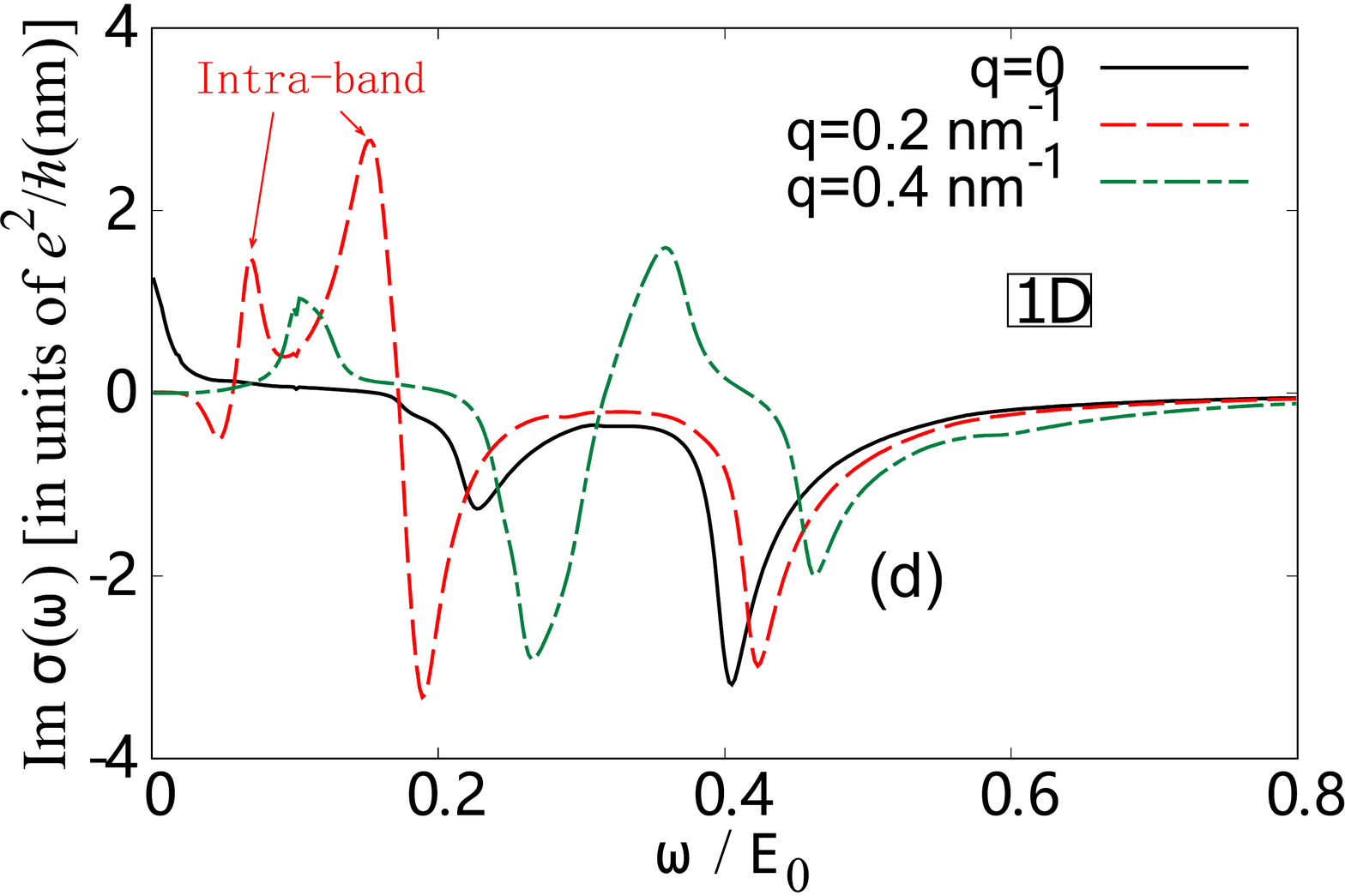} 

\end{centering}

\caption{(Color online) (a): the real part of the nonlocal optical conductivity for a 3D vector $\mathbf{q}$. The black solid curve is for the longitudinal conductivity $\sigma^L_{xx}(q,\omega)$ ($\mathbf{q}$ along $\hat{e}_{x}$), the Drude peak shifts to higher energy and splits into two peaks in the intraband, due to the two different Fermi velocities. The red dashed curve is for the transverse conductivity $\sigma^T_{xx}(q,\omega)$  ($\mathbf{q}$ perpendicular to $\hat{e}_{x}$), the Drude peak is not shifted.
In (b), two dimensional (2D) vector $\mathbf{q}$ is considered. We observe similar behavior as those in 3D. The second peak in the intraband is sharper than the 3D case. 
(c): The real part of the optical conductivity (black solid) for a doped CDW half-metal in 1D ($q=0.2$ $\mathrm{nm}^{-1}$) and the spin optical conductivity (red dashed) compared with the valley optical conductivity (green dash-dotted). It is clear that the two intra-band peaks are associated with the same direction of spin but with opposite valley. The two inter-band peaks are however associated with opposite spin. (d): the imaginary part of the nonlocal optical conductivity, crucial for plasmonics, two intra-band peaks in the imaginary part was confirmed for $q=0.2$ $\mathrm{nm}^{-1}$ (red dashed), as found in the Re$\sigma(q,\omega)$. In all the plots, the parameters ($\Delta_{\uparrow}=0.063 E_0$, $\Delta_{\downarrow}=0.2 E_0$) are the same for 1D, 2D and 3D. The chemical potential is $\mu=0.11 E_0$ intersecting with the lower conduction band. The temperature $T$=0.001 K. In all the plots of the Re$\sigma(q,\omega)$, the first two peaks (or one peak and one dip) in the region $0<\omega/E_0<0.2$ are considered as intra-band peaks, originating from two Fermi velocities. Then in the region $0.2<\omega/E_0$, we see two inter-band peaks, originating from the peaks in the joint density of states for Re$\sigma(q=0,\omega)$,\cite{Inter} or the joint spectral function for Re$\sigma(q\ne0,\omega)$.   }
\label{Fig5} 
\end{figure}

In 3D, we need to perform the four-fold integration (Eq.~(8)) numerically, which is very slow. For one frequency point it takes about 1 hour on 1 CPU. In Fig.~\ref{Fig5}(a) we use 1600 points and show the real part of the nonlocal optical conductivity in three dimensions for $q=0.2 \mathrm{nm}^{-1}$. We found the Drude peak at zero frequency is shifted to higher frequencies for the longitudinal conductivity. In the intra-band region ($0<\omega/E_0<0.2$), we found the Drude peak is separated into two-peak structure. This is due to the two Fermi velocities in the CDW as shown in Fig.~\ref{Fig4}. If the two typical Fermi velocities are $v_{F1}$ and $v_{F2}$, the two intra-band peaks are roughly at $\omega_1=q\times v_{F1}$ and $\omega_2=q\times v_{F2}$. For the transverse direction ($\mathbf{q}$ perpendicular to $\hat{e}_{x}$), the Drude peak is not shifted. In both cases, we found two inter-band peaks at around 0.2 and 0.4, due to the fact that for a four-band system we can define two sets of inter-band transitions. The inter-band peaks were shifted to higher energy as the transverse mode crossovers to the longitudinal mode. 

In Fig.~\ref{Fig5}(b), we investigate the 2D conductivity and found the spectral weight is redistributed between the intra-band and inter-band peaks. The intra-band peaks in 2D become sharper than those in 3D. In Fig.~\ref{Fig5}(c), we observe that the 1D intra-band  spin conductivity is always positive, while the intra-band valley conductivity oscillates from negative to positive. This suggests the charge carriers at the two Fermi momenta carry the same spin and opposite valley. 

Fig.~\ref{Fig5}(d) shows the 1D imaginary part of the nonlocal optical conductivity, which is connected to the real part through the Kramers-Kronig relation. At $q=0$, the imaginary part does not diverge as $1/\omega$, so $\omega \mathrm{Im} \sigma =0$. At $q=0.2$ $\mathrm{nm}^{-1}$, we see clearly the intra-band evolves into two positive oscillations. At much larger $q=0.4$ 
$\mathrm{nm}^{-1}$, however we only see one positive peak in the intra-band. This suggests that at small q, two plasmonic modes may be observed in the Electron Energy Loss Spectroscopy. The inter-band transitions in the spin susceptibility of the spin-valley half-metal was studied in \cite{Susc}. For the inter-band optical conductivity, it is connected to the joint density of states for Re$\sigma(q=0,\omega)$,\cite{Inter} or the joint spectral function for Re$\sigma(q\ne0,\omega)$. 

\begin{figure}[tp]
\begin{centering}
\includegraphics[width=3.2in,height=3.2in]{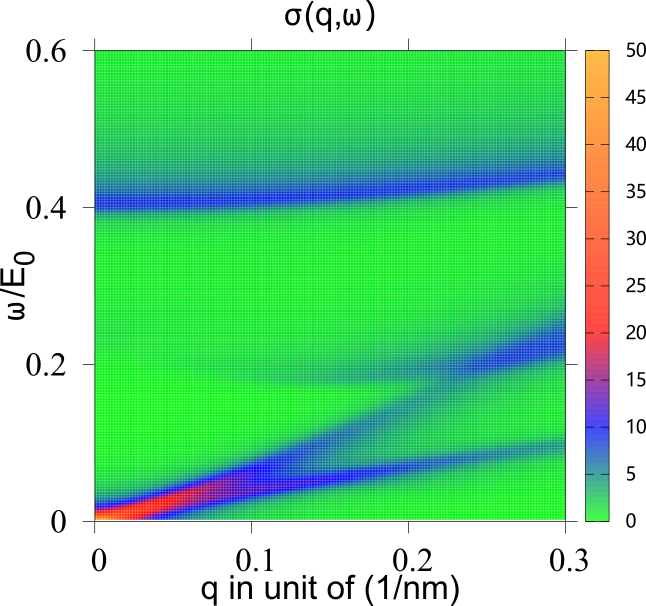} 

\par\end{centering}
\caption{(Color online) Real part of the nonlocal optical conductivity as a function of $\omega$ and $q$ in 1D.}
\label{Fig6} 
\end{figure}

In Fig.~\ref{Fig6}, we plot the real part of the nonlocal optical conductivity in 1D, we see the Drude peak at $q=0$ is shifted to higher energy, the spectral weight is transferred to the intra-band region. In the intra-band the Drude peak splits into two peaks, due to the two different Fermi velocities. The inter-band peaks at $\omega/E_0=0.4$ is not changed much by $q$. The parameters are the same as those in Fig.~\ref{Fig5}.

\begin{figure}[tp]
\begin{centering}
\includegraphics[width=3.2in,height=3in]{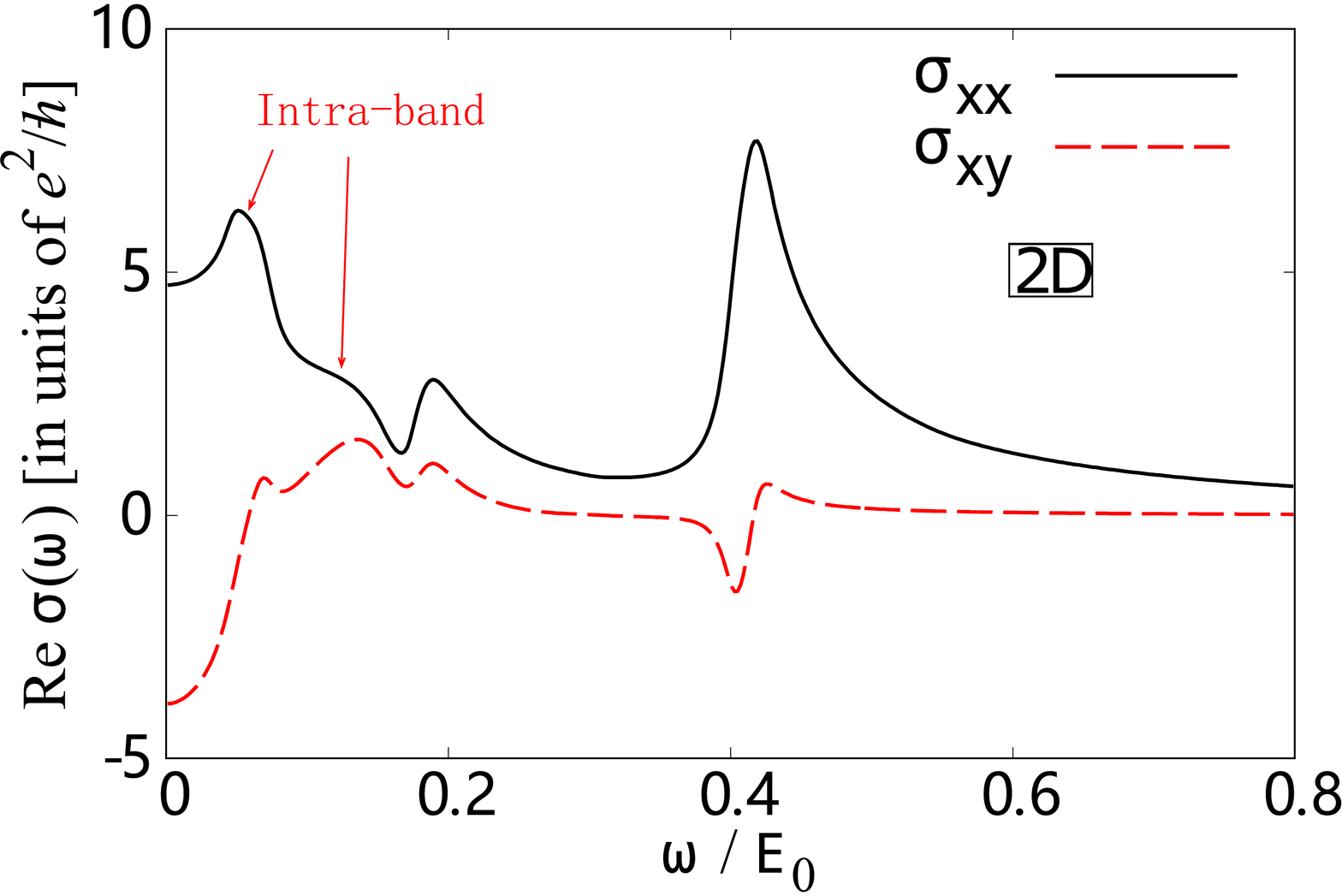} 

\par\end{centering}
\caption{(Color online) Real part of the nonlocal optical conductivity for a two dimension vector $\mathbf{q}$. The angle $\theta$ between the wave vector $\mathbf{q}$ and $\hat{e}_{x}$ is $\theta=\pi/4$, the wave vector is $q=0.2$ $\rm{nm}^{-1}$.The black solid curve is for the conductivity $\sigma_{xx}$, its behavior is in the transition from the longitudinal one $\sigma^L_{xx}$ to the transverse one $\sigma^T_{xx}$. The red dashed curve is for the off-diagonal conductivity $\sigma_{xy}$.}
\label{Fig7} 
\end{figure}

In Fig.~\ref{Fig7}, we investigate the off-diagonal conductivity $\sigma_{xy}(q,\omega)$, which is zero in the direction $\theta=0$ and $\theta=\pi/2$. Here we set the direction $\theta=\pi/4$ (between $\mathbf{q}$ and $\hat{e}_{x}$). We show that $\sigma_{xy}$ at $\omega=0$ is negative, and oscillates between positive and negative values at higher frequencies. The diagonal conductivity $\sigma_{xx}$ is in the transition from the transverse  ($\theta=\pi/2$) to the longitudinal ($\theta=0$) mode, we see a shifted Drude peak and a dip in the optical conductivity around $\omega/E_0=0.1$, due to the two Fermi velocities. 

\begin{figure}[tp]
\begin{centering}
\includegraphics[width=3.2in,height=3in]{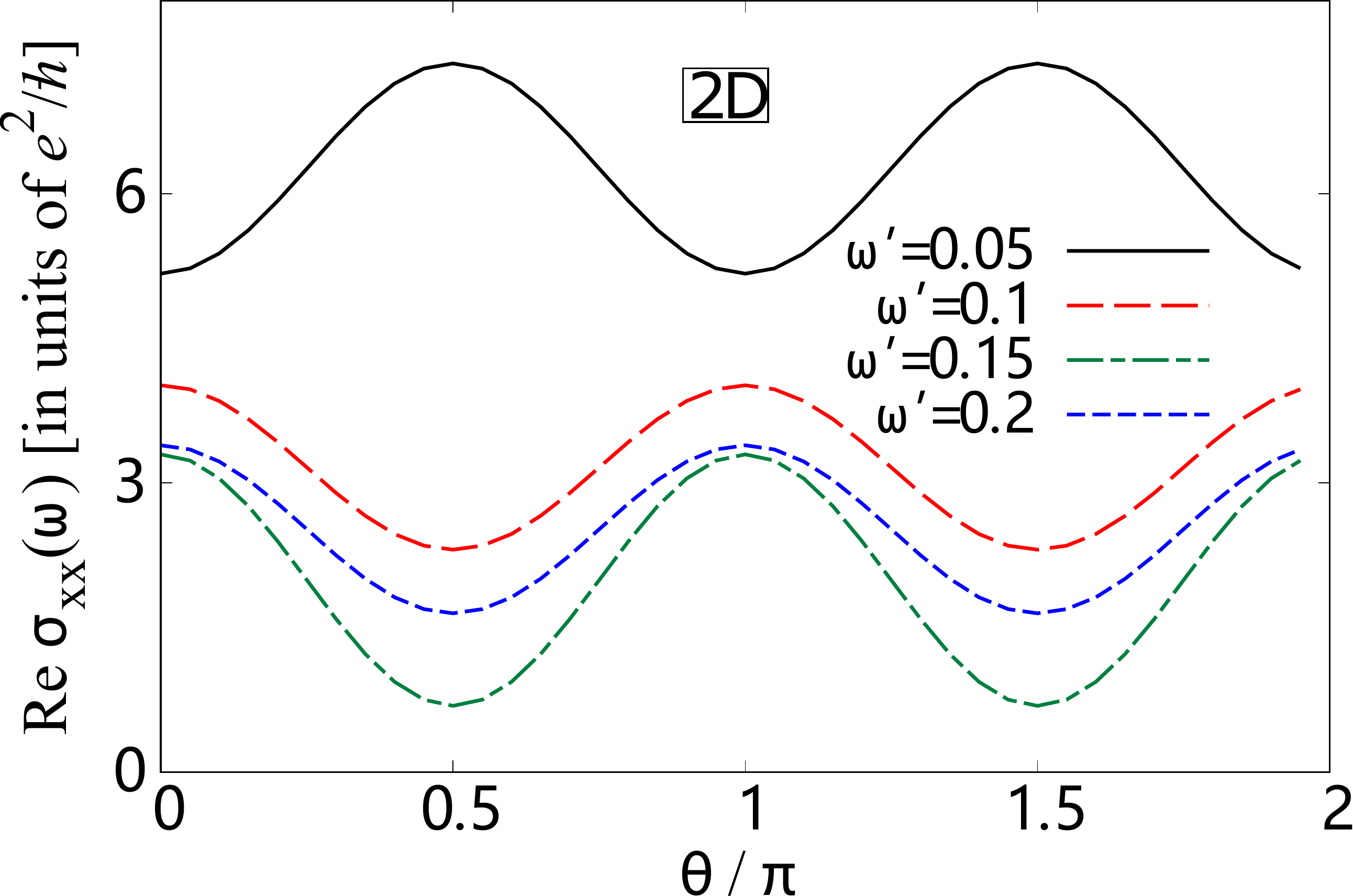} 
\par\end{centering}
\caption{(Color online) Real part of the nonlocal optical conductivity for a two dimension vector $\mathbf{q}$, as a function of the angle $\theta$ between the wave vector $\mathbf{q}$ and $\hat{e}_{x}$; the wave vector is $q=0.2$ $\rm{nm}^{-1}$. Here $\omega'=\omega/E_0$.}
\label{Fig8} 
\end{figure}

In Fig.~\ref{Fig8}, we observe that the $\sigma_{xx}(q,\omega)$ oscillates as the angle changes from  $\theta=0$ to $\theta=2\pi$, for four typical frequencies in the intra-band region. For three of them $\sigma_{xx}(q,\omega)$ reaches maximum at $\theta=0$, for one frequency at $\omega/E_0=0.05$, the maximum is at $\theta=\pi/2$. 

In table~IV, we present the DC off-diagonal conductivity along the direction $\theta=\pi/4$ for various wave vector q. We found the Im$\sigma_{xy}(q,0)$ is positive at small q, firstly increases then decreases and changes sign at larger q. In the supplementary material we present the numerical results of 1D optical conductivity and discussed the spin and valley conductivity respectively.
\begin{table}
\caption{\label{Sxy} Im$\sigma_{xy}(\mathbf{q},\omega=0)$ along the direction $\theta=\pi/4$.}
\begin{tabular}{c|c|c|c|c}
\hline 
\hline
 & $ q=0.001$  & $q=0.005$ & $q=0.01$ & $q=0.05$ \\
\hline 
Im$\sigma_{xy}(q,0)$ & 0.00025   & 0.0745  & -4.06 & -17.5 \\
\hline 
\end{tabular}
\end{table}

Note that in the Fermi surface nesting model, frequent electron-electron collisions modify the band-structure and two Fermi velocities appear. In a hydrodynamic approach, considering a fluid flow, the velocity changes gradually from the center to the boundary. In that sense, the CDW/SDW state lies somewhere between the ballistic transport picture and the fully hydrodynamic transport picture. The intra-band optical conductivity, which could be measured in a near-field experiment, is a good probe of these liquid-like properties. We also observe that the valley conductivity for one intra-band peak is negative, and for another intra-band peak is positive. This gives a clear definition of the valley. The positive valley is associated with the Fermi velocity $v_{F1}$ and the negative valley is associated with $v_{F2}$. In the above discussion we found the non-local intra-band optical conductivity is a useful tool to investigate the spin-valley physics.

\section{Conclusion}
In summary, we provide DFT calculations of Fermi surface nesting materials. Then based on the parabolic band approximation we derive the nonlocal optical conductivity for a spin/charge density wave Hamiltonian beyond the standard DFT. Our method is able to accurately capture the shift of the Drude peak from zero to higher frequency in the intra-band region and splits into two peaks, in 3D, 2D and 1D. As the direction changes from the transverse mode ($\theta=\pi/2$) to the longitudinal mode ($\theta=0$), although the Hamiltonian is isotropic, the nonlocal optical conductivity is anistropic at nonzero $q$. We discuss in detail the diagonal $\sigma_{xx}$ and off-diagonal $\sigma_{xy}$ conductivities at different angles.

\section{Methods}
A first-principles method based on the density-functional theory (DFT) is used. The Device Studio program provides a number of functions for performing visualization, modeling and simulation. We use the projector augmented wave method from the DS-PAW software integrated in the Device studio program. The Wannier fitting of DFT band structure is implemented within the WANNIER90 code. Many-body calculation based on the Green's function method is used.

\section{COMPETING INTERESTS}
The Authors declare no Competing Financial or Non-Financial Interests.

\section{DATA AVAILABILITY}
Data available on request from the authors. The data that support the findings of this study are available from the corresponding author, [Z.L.], upon reasonable request.

\section{AUTHOR CONTRIBUTIONS}
Z.L. designed the project and wrote the paper, X.H. carried out the DFT calculation, X.J. and B.H. provided the Wannier fitting, all authors contributed to the development of the work.

\begin{acknowledgments}
The authors thanks A. Rozhkov, A. Sboychakov, A. L. Rakhmanov, K. I. Kugel, Ryusuke Matsunaga and F. Nori for useful discussions. A part of this work has been supported by ISSP International Collaboration Program of the University of Tokyo. Z. L. acknowledges the support of a JSPS Foreign Postdoctoral Fellowship under Grant No. PE14052 and P16027 and the Chinese Academy of Science funding No. E1Z1D10200 and No. E2Z2D10200. This work is supported in part by the National Natural Science Foundation of China (Grant No. 61988102). We gratefully acknowledge HZWTECH for providing computation facilities.
\end{acknowledgments}

\onecolumngrid

\twocolumngrid

\appendix

\section{Derivations of the Green's function and the spectral function for a four-band Hamiltonian }

To obtain the Green's function associated with the Hamiltonian (1), we firstly use a general mathematical technique:  any 4 by 4 matrix can be partitioned into a smaller 2 by 2 block matrix. In the smaller block matrix, each element is a 2 by 2 sub-matrix. Use $A,B,C,D$ for the four sub-matrices, we have
\begin{eqnarray}
z-\hat{H}=\left[\begin{array}{cc}
A & C\\
D & B
\end{array}\right]
\end{eqnarray}
where $z=i\omega_n$ for imaginary frequency or $z=\omega+i\delta$ for real frequency. Here $A=(z-k^{2}/2m+\varepsilon_{0})\hat{I_{2}}$,
$B=(z+k^{2}/2m-\varepsilon_{0})\hat{I_{2}}$ and 
\begin{eqnarray}
C=\left[\begin{array}{cc}
0 & -\Delta_{\downarrow}\\
-\Delta_{\uparrow} & 0
\end{array}\right],D=\left[\begin{array}{cc}
0 & -\Delta_{\uparrow}\\
-\Delta_{\downarrow} & 0
\end{array}\right]
\end{eqnarray}
The Green's function matrix is also partitioned into a smaller 2 by 2 block matrix, with $G_{ee},G_{eh},G_{he},G_{hh}$ the four sub-matrix elements. The equation to determine the Green's function is written as
\begin{eqnarray}
 &  & \left[\begin{array}{cc}
A & C\\
D & B
\end{array}\right]\left[\begin{array}{cc}
G_{ee} & G_{eh}\\
G_{he} & G_{hh}
\end{array}\right]=\left[\begin{array}{cc}
\hat{I_{2}} & 0\\
0 & \hat{I_{2}}
\end{array}\right]
\end{eqnarray}
The sub-matrix elements $G_{ee}$ and $G_{he}$ satisfy the following equations,
\begin{eqnarray}
 & A\times G_{ee}+C\times G_{he}=\hat{I_{2}}
\end{eqnarray}
\begin{eqnarray}
 & B\times G_{he}+D\times G_{ee}=0
\end{eqnarray}
From the two equations we obtain
\begin{eqnarray}
 (AB-CD)G_{ee}=B
\end{eqnarray}
\begin{eqnarray}
 (AB-DC)G_{he}=-D
\end{eqnarray}
The product of  $CD$ is given by,
\begin{eqnarray}
\left[\begin{array}{cc}
0 & -\Delta_{\downarrow}\\
-\Delta_{\uparrow} & 0
\end{array}\right]\left[\begin{array}{cc}
0 & -\Delta_{\uparrow}\\
-\Delta_{\downarrow} & 0
\end{array}\right]=\left[\begin{array}{cc}
\Delta_{\downarrow}^{2} & 0\\
0 & \Delta_{\uparrow}^{2}
\end{array}\right]
\end{eqnarray}
For the other two sub-matrix elements $G_{eh}$ and $G_{hh}$, We
need to solve the two equations below 
\begin{eqnarray}
 A\times G_{eh}+C\times G_{hh}=0
\end{eqnarray}
\begin{eqnarray}
 D\times G_{eh}+B\times G_{hh}=\hat{I_{2}}
\end{eqnarray}
from which we obtained
\begin{eqnarray}
 (AB-CD)G_{eh}=-C
\end{eqnarray}
\begin{eqnarray}
(AB-DC)G_{hh}=A
\end{eqnarray}
Notice the poles of the four Green's functions are given by the equation 
\begin{equation}
(z-k^{2}/2m+\varepsilon_{0})(z+k^{2}/2m-\varepsilon_{0})-\Delta^{2}=0
\end{equation}
where $\Delta=\Delta_{\uparrow}$ or $\Delta_{\downarrow}$,
expand this equation in orders of $z$ we have, 
\begin{equation}
z^{2}-(k^{2}/2m-\varepsilon_{0})^2-\Delta^{2}=0
\end{equation}
From which we obtain two group of solutions for the eigenvalues corresponding to $\Delta=\Delta_{\uparrow}$ and $\Delta_{\downarrow}$ separately,
\begin{equation}
E_{1,2}=\pm\sqrt{(\frac{k^{2}}{2m}-\varepsilon_{0})^2+\Delta_{\uparrow}^{2}}
\end{equation}
and the other two for $\Delta_{\downarrow}$,
\begin{equation}
E_{3,4}=\pm\sqrt{(\frac{k^{2}}{2m}-\varepsilon_{0})^2+\Delta_{\downarrow}^{2}}
\end{equation}
With the eigenvalues $E_{1,2,3,4}$ we can write the Green's function in a compact form
\begin{equation}
G_{ee}=\left[\begin{array}{cc}
\frac{(z+k^{2}/2m-\varepsilon_{0})}{(z-E_{3})(z-E_{4})} & 0\\
0 & \frac{(z+k^{2}/2m-\varepsilon_{0})}{(z-E_{1})(z-E_{2})}
\end{array}\right]
\end{equation}
\begin{equation}
G_{he}=\left[\begin{array}{cc}
0 & \frac{\Delta_{\uparrow}}{(z-E_{1})(z-E_{2})}\\
\frac{\Delta_{\downarrow}}{(z-E_{3})(z-E_{4})} & 0
\end{array}\right]
\end{equation}
\begin{equation}
 G_{hh}=\left[\begin{array}{cc}
\frac{(z-k^{2}/2m+\varepsilon_{0})}{(z-E_{1})(z-E_{2})} & 0\\
0 & \frac{(z-k^{2}/2m+\varepsilon_{0})}{(z-E_{3})(z-E_{4})}
\end{array}\right]
\end{equation}
\begin{equation}
G_{eh}=\left[\begin{array}{cc}
0 & \frac{\Delta_{\downarrow}}{(z-E_{3})(z-E_{4})}\\
\frac{\Delta_{\uparrow}}{(z-E_{1})(z-E_{2})} & 0
\end{array}\right]
\end{equation}
To obtain the spectral function $\hat{A}(\mathbf{k},\omega)$, which is the imaginary part of the Green's function, we need to use the following rules to decompose the Green's function into a sum of single poles,
\begin{eqnarray}
&& \frac{(z+k^{2}/2m-\varepsilon_{0})}{(z-E_{1})(z-E_{2})}=\frac{1}{(E_{1}-E_{2})}\times \nonumber\\
&& \Big[\frac{E_{1}+k^{2}/2m-\varepsilon_{0}}{(z-E_{1})}-\frac{E_{2}+k^{2}/2m-\varepsilon_{0}}{(z-E_{2})}\Big]
\end{eqnarray}
\begin{eqnarray}
&& \frac{(z-k^{2}/2m+\varepsilon_{0})}{(z-E_{1})(z-E_{2})}=\frac{1}{(E_{1}-E_{2})}\times \nonumber\\
&& \Big[\frac{E_{1}-k^{2}/2m+\varepsilon_{0}}{(z-E_{1})}-\frac{E_{2}-k^{2}/2m+\varepsilon_{0}}{(z-E_{2})}\Big]
\end{eqnarray}
\begin{equation}
\frac{\Delta_{\uparrow}}{(z-E_{1})(z-E_{2})}=\frac{\Delta_{\uparrow}}{(E_{1}-E_{2})}\Big[\frac{1}{(z-E_{1})}-\frac{1}{(z-E_{2})}\Big]
\end{equation}
The density of states (DOS) $N(\omega)$ can be obtained through the
spectral function: $N(\omega)=\sum_{\mathbf{k}}\mathrm{Tr}[\hat{A}(\mathbf{k},\omega)]$.

\section{Derivation of the nonlocal optical conductivity }
In order to evaluate the conductivity we need to take the
trace of the matrix product of velocity operator and Green's function
matrix, for $\sigma_{xx}(\omega)$ we need $\mathrm{Tr}\langle\hat{v}_{x}\widehat{G}(\mathbf{k}_{1},\omega_{1})\hat{v}_{x}\widehat{G}(\mathbf{k}_{2},\omega_{2})\rangle=(k_x/m)^{2}\times$
\begin{eqnarray}
&& \{\mathrm{Tr}[G_{ee}(\mathbf{k}_{1},\omega_{1})G_{ee}(\mathbf{k}_{2},\omega_{2})]\nonumber\\
&& -\mathrm{Tr}[G_{eh}(\mathbf{k}_{1},\omega_{1})G_{he}(\mathbf{k}_{2},\omega_{2})+G_{he}(\mathbf{k}_{1},\omega_{1})G_{eh}(\mathbf{k}_{2},\omega_{2})]\nonumber\\
&& +\mathrm{Tr}[G_{hh}(\mathbf{k}_{1},\omega_{1})G_{hh}(\mathbf{k}_{2},\omega_{2})]\}
\end{eqnarray}
The spin and valley conductivity are given as $\mathrm{Re}\sigma^{s,v}_{xx}(\mathbf{q},\omega)=\frac{e^{2}}{4\pi\omega}\int_{-\infty}^{\infty}d\omega_{1}[f(\omega_{1})-f(\omega_{1}+\omega)]\times$
\begin{equation*}
\sum_{\mathbf{k}}\mathrm{Tr}\langle S_z\hat{v}_{x}\hat{A}(\mathbf{k}-\mathbf{q}/2,\omega_{1})\hat{v}_{x}\hat{A}(\mathbf{k}+\mathbf{q}/2,\omega_{1}+\omega)\rangle
\end{equation*}
and
\begin{equation*}
\sum_{\mathbf{k}}\mathrm{Tr}\langle \tau_z\hat{v}_{x}\hat{A}(\mathbf{k}-\mathbf{q}/2,\omega_{1})\hat{v}_{x}\hat{A}(\mathbf{k}+\mathbf{q}/2,\omega_{1}+\omega)\rangle
\end{equation*}
respectively.

\begin{figure}[tp]
\begin{centering}
\includegraphics[width=3.2in,height=2.8in]{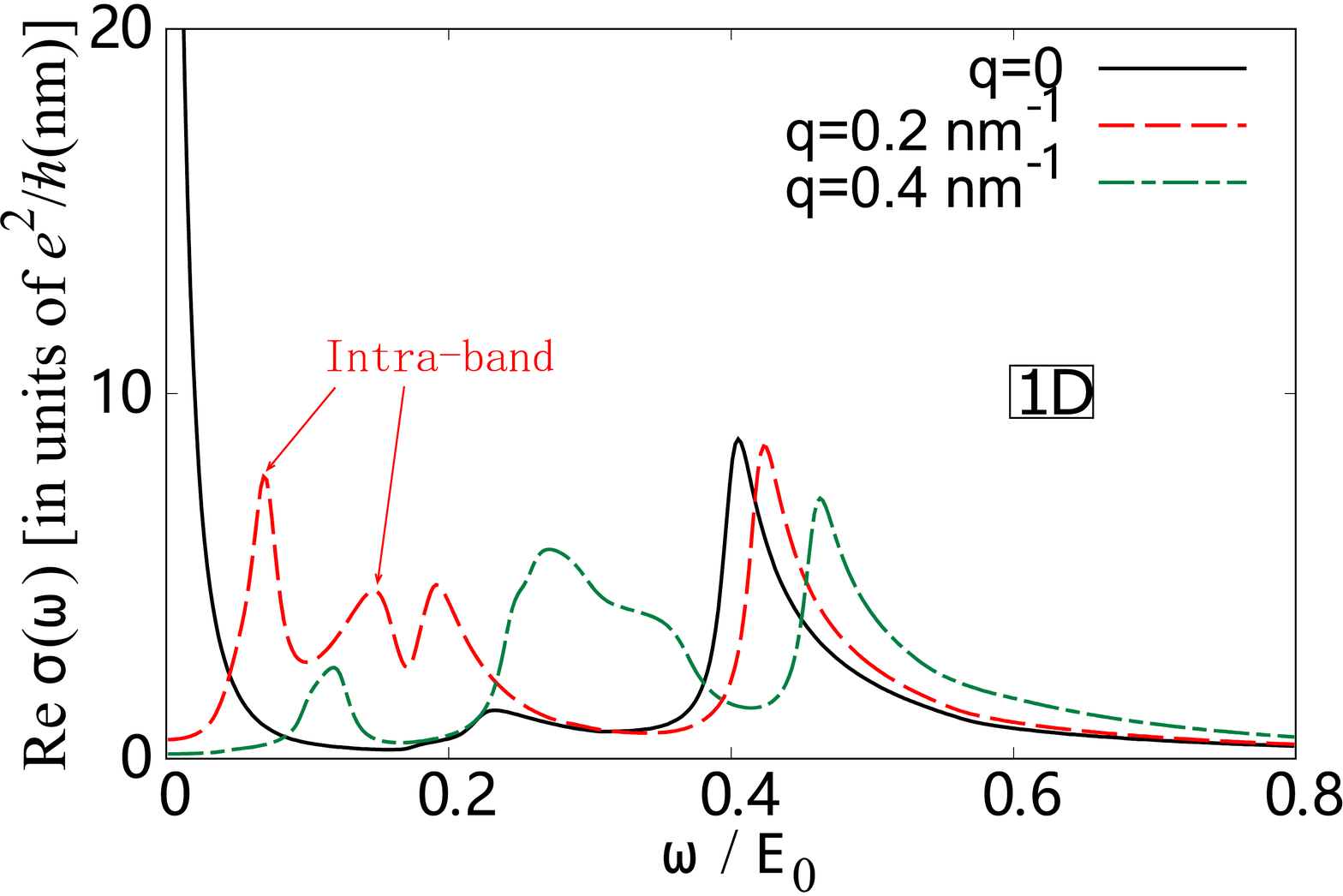} 
\par\end{centering}
\caption{(Color online) Real part of the optical conductivity for $\Delta_{\uparrow}=0.063 E_0$ and $\Delta_{\downarrow}=0.2 E_0$, for wave vectors $q=0$, $q=0.2$ $\mathrm{nm}^{-1}$ and $q=0.4$ $\mathrm{nm}^{-1}$. The chemical potential $\mu=0.11 E_0$. For $q=0$ (black solid), a Drude peak from intra-band transition is found near zero frequency, at higher frequencies $\omega=0.2 E_0$ and $\omega=0.4 E_0$ two peaks from inter-band transition are found, due to the four-band feature of the Hamiltonian. For $q=0.2$ $\mathrm{nm}^{-1}$ (red dashed), the Drude peak is shifted to higher energy and separated into two peaks in the frequency region $\omega<0.2 E_0$. For $q=0.4$ $\mathrm{nm}^{-1}$ (green dash-dotted), only one peak in the frequency region $\omega<0.2 E_0$ and the other peak shifts to the inter-band region $\omega>0.2 E_0$. The inter-band  peaks should be compared with the peaks in the joint density of states for Re$\sigma(q=0,\omega)$, or the joint spectral function for Re$\sigma(q\ne0,\omega)$. }
\label{figA1} 
\end{figure}

For the charge density wave, we define the spin conductivity $\sigma^{s}_{xx}(\omega)$, the spin operator is $S_z=\mathrm{diag}(-1,1,1,-1)$; for the spin density wave, the spin operator becomes $S_z=\mathrm{diag}(-1,1,-1,1)$, the spin valley operator is $S_z\tau_z=\mathrm{diag}(-1,1,1,-1)$, the same as the spin operator for the charge density wave. Mathematically, for the spin conductivity $\sigma^{s}_{xx}(\omega)$ (charge density wave) and the spin valley conductivity $\sigma^{sv}_{xx}(\omega)$ (spin density wave), we need the same trace, $\mathrm{Tr}\langle S_z \hat{v}_{x}\widehat{G}(\mathbf{k}_{1},\omega_{1})\hat{v}_{x}\widehat{G}(\mathbf{k}_{2},\omega_{2})\rangle=(k_x/m)^{2}\times$
\begin{eqnarray}
&& \{-\mathrm{Tr}[\sigma_z G_{ee}(\mathbf{k}_{1},\omega_{1})G_{ee}(\mathbf{k}_{2},\omega_{2})]\nonumber\\
&& -\mathrm{Tr}[\sigma_z (G_{he}(\mathbf{k}_{1},\omega_{1})G_{eh}(\mathbf{k}_{2},\omega_{2})-G_{eh}(\mathbf{k}_{1},\omega_{1})G_{he}(\mathbf{k}_{2},\omega_{2}))]\nonumber\\
&& +\mathrm{Tr}[\sigma_z G_{hh}(\mathbf{k}_{1},\omega_{1})G_{hh}(\mathbf{k}_{2},\omega_{2})]\}
\end{eqnarray}
For the valley conductivity $\sigma^{v}_{xx}(\omega)$ we need $\mathrm{Tr}\langle \tau_z \hat{v}_{x}\widehat{G}(\mathbf{k}_{1},\omega_{1})\hat{v}_{x}\widehat{G}(\mathbf{k}_{2},\omega_{2})\rangle=(k_x/m)^{2}\times$
\begin{eqnarray}
&& \{\mathrm{Tr}[G_{ee}(\mathbf{k}_{1},\omega_{1})G_{ee}(\mathbf{k}_{2},\omega_{2})]\nonumber\\
&& -\mathrm{Tr}[G_{eh}(\mathbf{k}_{1},\omega_{1})G_{he}(\mathbf{k}_{2},\omega_{2})-G_{he}(\mathbf{k}_{1},\omega_{1})G_{eh}(\mathbf{k}_{2},\omega_{2})]\nonumber\\
&& -\mathrm{Tr}[G_{hh}(\mathbf{k}_{1},\omega_{1})G_{hh}(\mathbf{k}_{2},\omega_{2})]\}
\end{eqnarray}

In the clean limit where the impurity-scattering self-energy vanishes, the spectral function reduces to $\delta$-function, for example, $A_{ee}(\mathbf{k},\omega)=$
\begin{equation}
-2\pi \left[\begin{array}{cc}
\frac{E'_3\delta(\omega-E_{3})-E'_4\delta(\omega-E_{4})}{(E_{3}-E_{4})} & 0\\
0 & \frac{E'_1\delta(\omega-E_{1})-E'_2\delta(\omega-E_{2})}{(E_{1}-E_{2})}
\end{array}\right]
\end{equation}
where $E'_n=E_{n}+k^{2}/2m-\varepsilon_{0}$ and $n=1,2,3,4$. The trace $\mathrm{Tr}[G_{ee}(\mathbf{k}_{1},\omega_{1})G_{ee}(\mathbf{k}_{2},\omega_{2})]=$
\begin{eqnarray*}
G_{ee}(\mathbf{k}_{1},\omega_{1},\downarrow)G_{ee}(\mathbf{k}_{2},\omega_{2},\downarrow)+G_{ee}(\mathbf{k}_{1},\omega_{1},\uparrow)G_{ee}(\mathbf{k}_{2},\omega_{2},\uparrow)
\end{eqnarray*}
where $G_{ee}(\mathbf{k}_1,\omega_{1},\uparrow)=\frac{(z_{1}+k_1^{2}/2m-\varepsilon_{0})}{(z_{1}-E_{1})(z_{1}-E_{2})}$
and $G_{ee}(\mathbf{k}_{1},\omega_{1},\downarrow)=\frac{(z_{1}+k_1^{2}/2m-\varepsilon_{0})}{(z_{1}-E_{3})(z_{1}-E_{4})}$.

In 1D, the nonlocal optical conductivity contains the term $\frac{E'_1(k-q/2)E'_1(k+q/2)}{E_{1}(k-q/2)E_{1}(k+q/2)}\delta(\omega_1-E_1(k-q/2))\delta(\omega_1+\omega-E_1(k+q/2))$, the two $\delta$-functions could be used to carry out the integral over $\omega_1$, and we are left with the one dimension integral over $k$. The upper and lower limit in the integral over $k$ is set by the Fermi-Dirac distribution function. The analytical result will be very long and complicated because many terms from intra-band and inter-band need to be considered. Instead we decide to carry out the integration numerically. The self-energy from impurity scattering is nonzero in all the numerical results.

In Fig.~\ref{figA1}, we show the real part of the optical conductivity for various wave vectors $q$. For $q=0$, we recover the famous Drude peak for the intra-band optical conductivity. For non-zero wave-vector $q$, we observe the shift of the Drude peak to higher frequency. The single Drude peak is separated into two peaks, which are associated with the two typical energies for the intra-band transition. This is due to the fact that in one dimension, the slope near the four Fermi momenta (the momenta at which the Fermi surface intersects with the band) are different. The slope is related to the Fermi velocity. There are two typical Fermi velocities $v_{F1}$ and $v_{F2}$. Thus for the same $q$ we observe two intra-band peaks roughly at $\omega_1=q\times v_{F1}$ and $\omega_2=q\times v_{F2}$, in the frequency region $\omega<0.2 E_0$. The inter-band region starts from around $\omega>0.2 E_0$, as observed in the optical conductivity for the case $q=0$. The peaks in the inter-band region are determined from the peaks in the joint density of states. For the cases  $q\ne0$, the peaks in the inter-band region are determined from the peaks in the joint spectral functions.

\section{Band structure and a tight binding model of monolayer $\rm{VSe_2}$}

\begin{figure}[tp]
    \begin{centering}
    \includegraphics[width=3.5in]{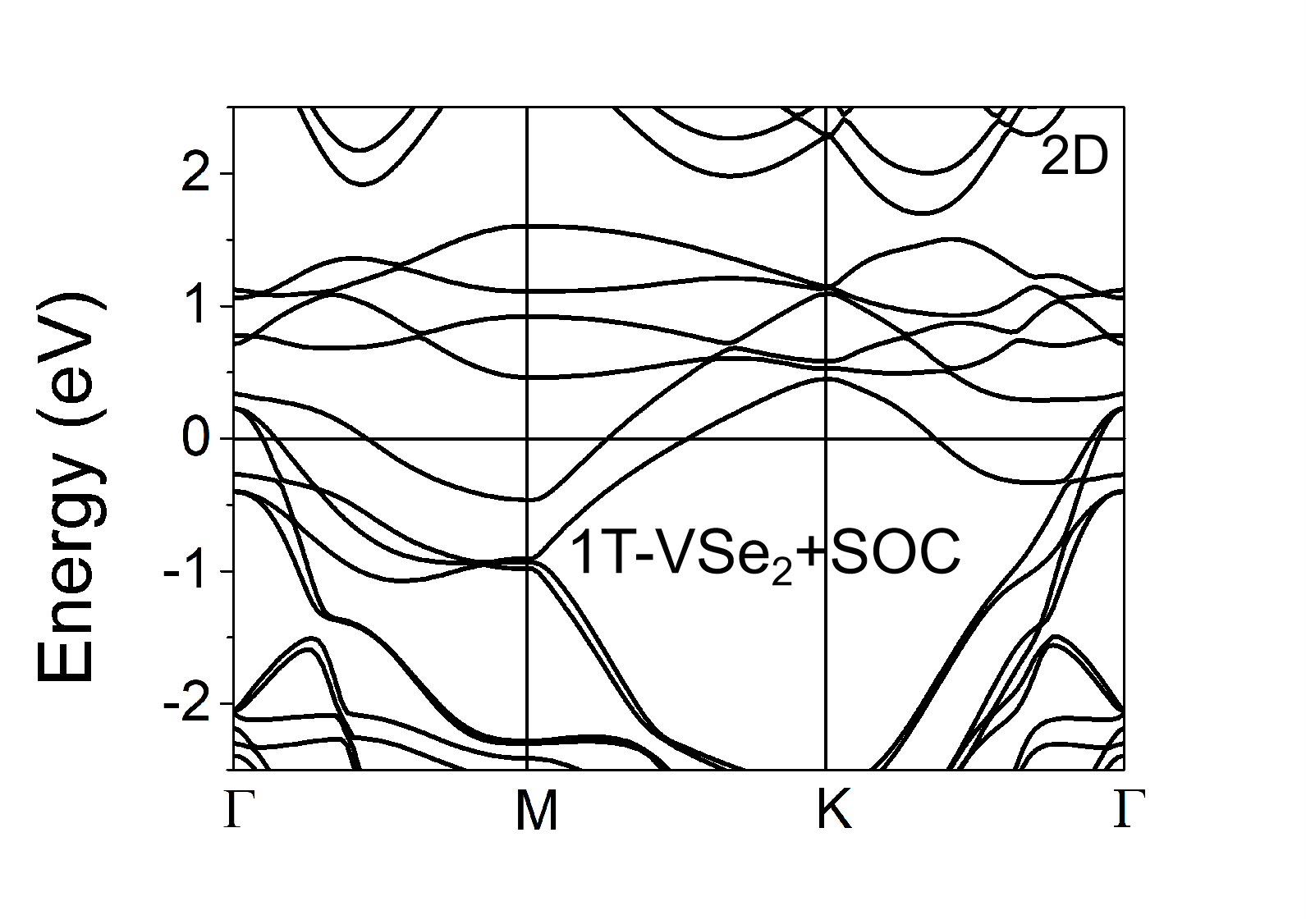}
    \par\end{centering}
    \caption{The band structure of monolayer $\rm{VSe_2}$ from DFT calculation with spin orbit coupling (SOC).Note that the band structure is close to the spin polarized DFT calculation as in the Fig.~2(a) of the main text. In the 2D calculation, we set the lattice constant $c=20 \AA$ to ensure that there is no interaction between the layers, to achieve an ideal single-layer 2D material condition.}
    \label{fig:soc}
\end{figure}

\begin{figure}[tp]
    \begin{centering}
    \includegraphics[width=3.5in]{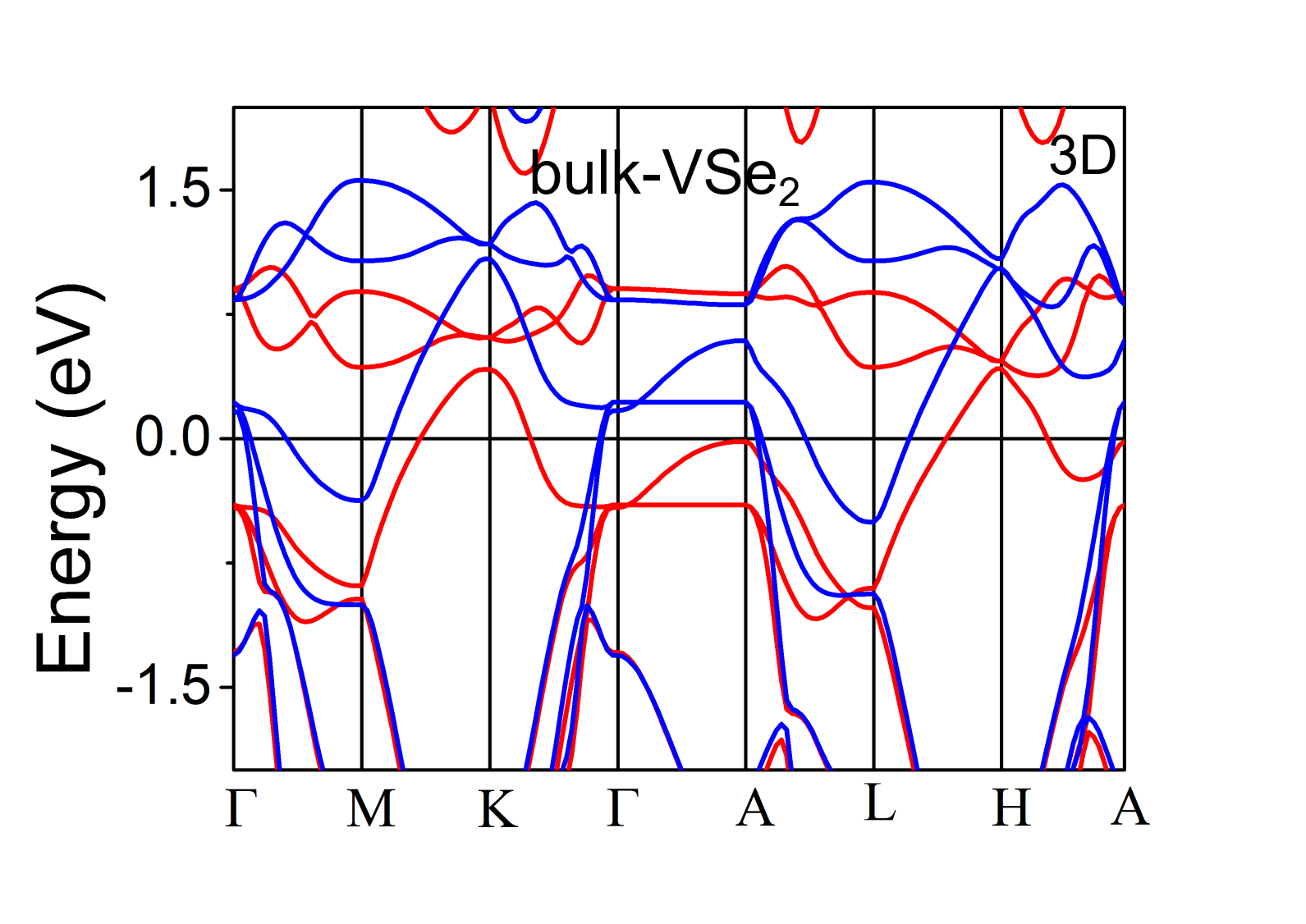}
    \par\end{centering}
    \caption{(Color online) The band structure of 3D $\rm{VSe_2}$ from spin polarized DFT calculation. Note that the band structure around the points $\Gamma$, $M$, $K$ is close to those in the 2D monolayer $\rm{VSe_2}$ (Fig.~2(a) of the main text). In 3D we have more high symmetry momentum points $H$, $A$, $L$ and observe different band structure around these points as expected.  }
    \label{fig:bulk}
\end{figure}

\begin{figure}[tp]
    \begin{centering}
    \includegraphics[width=3.5in]{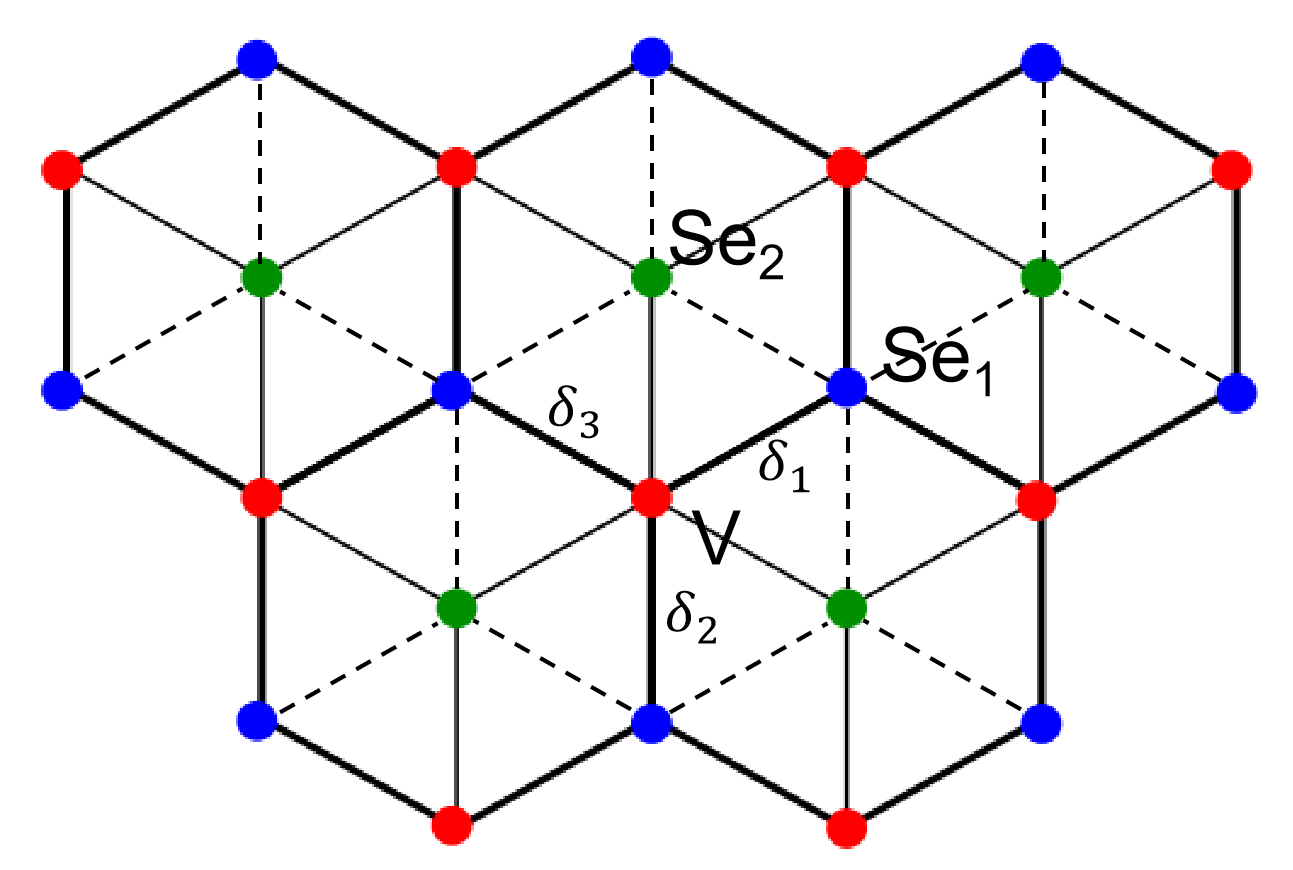}
    \par\end{centering}
    \caption{(Color online) A simplified tight binding model based on the Wannier fitting of the band structure of monolayer $\rm{VSe_2}$. Solid line represent the hopping between V and Se atoms, the hopping parameters are given in table II of the main text. Dashed line represent the hopping between $\rm{Se_1}$ and $\rm{Se_2}$ atoms, the hopping parameters are given in table I of the main text. Note that the dashed line hopping terms are not included in the approximate tight binding model $H_t$. }
    \label{fig:TB}
\end{figure}

\begin{table}[]
\caption{ The V-V hopping amplitude.}
\begin{tabular}{|cc|ccccc|}
\hline
\multicolumn{2}{|c|}{\multirow{2}{*}{}} & \multicolumn{5}{c|}{\textbf{V(1 0 0)}} \\ \cline{3-7} 
\multicolumn{2}{|c|}{} & \multicolumn{1}{c|}{$d_{xy}$} & \multicolumn{1}{c|}{$d_{yz}$} & \multicolumn{1}{c|}{$d_{z^2}$} & \multicolumn{1}{c|}{$d_{xz}$} & $d_{x^2-y^2}$ \\ \hline
\multicolumn{1}{|c|}{\multirow{5}{*}{\textbf{V}}} & $d_{xy}$ & \multicolumn{1}{c|}{-0.183} & \multicolumn{1}{c|}{0} & \multicolumn{1}{c|}{-0.052} & \multicolumn{1}{c|}{0.153} & 0 \\ \cline{2-7} 
\multicolumn{1}{|c|}{} & $d_{yz}$ & \multicolumn{1}{c|}{0} & \multicolumn{1}{c|}{-0.052} & \multicolumn{1}{c|}{0} & \multicolumn{1}{c|}{0} & 0.215 \\ \cline{2-7} 
\multicolumn{1}{|c|}{} & $d_{z^2}$ & \multicolumn{1}{c|}{-0.052} & \multicolumn{1}{c|}{0} & \multicolumn{1}{c|}{0.095} & \multicolumn{1}{c|}{-0.085} & 0 \\ \cline{2-7} 
\multicolumn{1}{|c|}{} & $d_{xz}$ & \multicolumn{1}{c|}{0.153} & \multicolumn{1}{c|}{0} & \multicolumn{1}{c|}{-0.085} & \multicolumn{1}{c|}{-0.271} & 0 \\ \cline{2-7} 
\multicolumn{1}{|c|}{} & $d_{x^2-y^2}$ & \multicolumn{1}{c|}{0} & \multicolumn{1}{c|}{0.215} & \multicolumn{1}{c|}{0} & \multicolumn{1}{c|}{0} & 0.135 \\ \hline
\end{tabular}
\label{table:V-V}
\end{table}

In this section, we present DFT calculations of the 2D and 3D $\rm{VSe_2}$ band structure to be compared with the results presented in the main text. In Fig.~\ref{fig:soc}, we present the DFT calculation with the spin-orbit coupling (SOC) effect and observe the impact of SOC on 1T-$\rm{VSe_2}$ is minimal. In Fig.~\ref{fig:bulk}, we present the 3D $\rm{VSe_2}$ band structure.

A simplified tight binding model could be obtained by using the hopping parameters in the table I, II and III of the main text. In Fig.~\ref{fig:TB}, we use a schematic to describe the hoppings on the lattice. The solid line represent the hopping between V and Se atoms, the hopping parameters are given in table II of the main text, the model is given below
\begin{eqnarray}
&& H_t=\sum_{i,\delta,d,p}[t_{dp1}(|V_{i,d}><Se_{1,i,\delta,p}|+|Se_{1,i,\delta,p}><V_{i,d}|)+\nonumber\\
&& t_{dp2}(|V_{i,d}><Se_{2,i,-\delta,p}|+|Se_{2,i,-\delta,p}><V_{i,d}|)]
\end{eqnarray}
where $i$ is the lattice site, $\delta=\delta_{1,2,3}$ are the three hopping vectors, $d$ are the five $d$ orbitals of V atom, $p$ are the three $p$ orbitals of Se atom, $t_{dp1}$ and $t_{dp2}$ are the hopping parameters given in table II of the main text. Here $\delta_1=(1.667,0.962)$, $\delta_2=(0.000,-1.924)$ and $\delta_3=(-1.667,0.962)$ in unit of $\AA$. The tight binding Hamiltonian is a 11 by 11 matrix. The dashed line in Fig.~\ref{fig:TB} represents the hopping between $\rm{Se_1}$ and $\rm{Se_2}$ atoms, the hopping parameters are given in table I of the main text. The next nearest neighbor hopping between Se atoms could be obtained from table III of the main text. In table I of this supplementary material, we list the nearest neighbor V-V hopping parameters for $\rm{VSe_2}$ from the cell (000) to (100), which is much smaller than the other hopping parameters.

\end{document}